\documentclass[10pt,conference,letterpaper]{IEEEtran}
\usepackage{booktabs} 
\usepackage{amsmath,url,graphicx,amssymb}
\usepackage{amsfonts}
\usepackage{ctable}
\usepackage{multirow}
\usepackage{algorithm}
\usepackage{algpseudocode}
\usepackage{pifont}
\usepackage{color}
\usepackage{subfigure}
\usepackage{enumitem}

\newcommand{\mylistbegin}{
  \begin{list}{$\bullet$}
   {
     \setlength{\itemsep}{-2pt}
     \setlength{\leftmargin}{1em}
     \setlength{\labelwidth}{1em}
     \setlength{\labelsep}{0.5em} } }
\newcommand{\mylistend}{
   \end{list}  }

\newcommand{\eg}{\textit{e.g.}}
\newcommand{\xeg}{\textit{E.g.}}
\newcommand{\ie}{\textit{i.e.}}
\newcommand{\etc}{\textit{etc}}

\newcommand{\wrt}{\textit{w.r.t.~}}

\title{CONE: Community Oriented Network Embedding}
\author{
{Carl Yang{\small $~^{\#}$}, Hanqing Lu{\small $~^{*}$}, Kevin Chen-Chuan Chang{\small $~^{\#}$} }
\vspace{1.2mm}\\
\fontsize{10}{10}\selectfont\itshape
$^{\#}$\,University of Illinois, Urbana Champaign\\
201 N Goodwin Ave, Urbana, Illinois 61801, USA\\
\fontsize{9}{9}\selectfont\ttfamily\upshape
\{jiyang3, kcchang\}@illinois.edu
\vspace{1mm}\\
\fontsize{10}{10}\selectfont\rmfamily\itshape
$^{*}$\,Carnegie Mellon University\\
5000 Forbes Ave, Pittsburgh, Pennsylvania  15213, USA\\
\fontsize{9}{9}\selectfont\ttfamily\upshape
hanqinglu@cmu.edu
}

\begin{document}
\maketitle

\setlength{\floatsep}{4pt plus 4pt minus 1pt}
\setlength{\textfloatsep}{4pt plus 2pt minus 2pt}
\setlength{\intextsep}{4pt plus 2pt minus 2pt}

\setlength{\dbltextfloatsep}{3pt plus 2pt minus 1pt}
\setlength{\dblfloatsep}{3pt plus 2pt minus 1pt}
\setlength{\abovecaptionskip}{3pt}
\setlength{\belowcaptionskip}{2pt}
\setlength{\abovedisplayskip}{2pt plus 1pt minus 1pt}
\setlength{\belowdisplayskip}{2pt plus 1pt minus 1pt}

\begin{abstract}
Detecting communities has long been popular in the research on networks. It is usually modeled as an unsupervised clustering problem on graphs, based on heuristic assumptions about community characteristics, such as edge density and node homogeneity. In this work, we doubt the universality of these widely adopted assumptions and compare human labeled communities with machine predicted ones obtained via typical mainstream algorithms. Based on supportive results, we argue that communities are defined by their underlying social patterns and unsupervised learning algorithms based on heuristics is incapable of capturing their various forms. Therefore, we propose to inject supervision into community detection through Community Oriented Network Embedding (CONE), which leverages limited ground-truth communities as examples to learn an embedding model aware of the underlying social patterns. Specifically, a deep architecture is developed by combining recurrent neural networks with random-walks on graphs towards capturing social patterns directed by ground-truth communities. Generic clustering algorithms on the embeddings of other nodes produced by the learned model then effectively reveals more communities that share similar social patterns with the ground-truth ones. 
\end{abstract}


\section{Introduction}
\label{sec:intro}
One of the most popular topics in network research is to identify communities. 
On the one hand, as networks are growing larger than ever before, it is efficient and necessary to look into smaller sub-networks, which consist of specific groups of interacting objects (\ie, nodes) and their links (\ie, edges). 
On the other hand, the knowledge of community structures allows us to better understand the status of an object within a group and the relations between it and its peers, so as to enable multiple benefits including the discovery of functionally related objects \cite{streich2009multi}, the study of interactions between modules \cite{airoldi2008mixed}, the inference of missing contents \cite{li2014user}, the prediction of unobserved connections \cite{chang2009relational} and so on. 

Many algorithms aim to solve this problem \cite{akoglu2012pics,chakraborty2015formation, clauset2004finding, mcauley2012learning,yang2013overlapping, yang2013community, zhou2012community}. However, they are all formulated based on the heuristic assumptions of edge density and node homogeneity, \ie, communities are constructed by densely connected nodes and nodes within one community are all homogeneous in some ways. Most surprisingly, even the ground-truth community labels used for evaluation in many standard datasets are generated by machines rather than humans based on these two assumptions \cite{yang2015defining}. 
In this work, we doubt the universality of these widely trusted assumptions.

\begin{figure}[]
\centering
\includegraphics[width=0.4\textwidth]{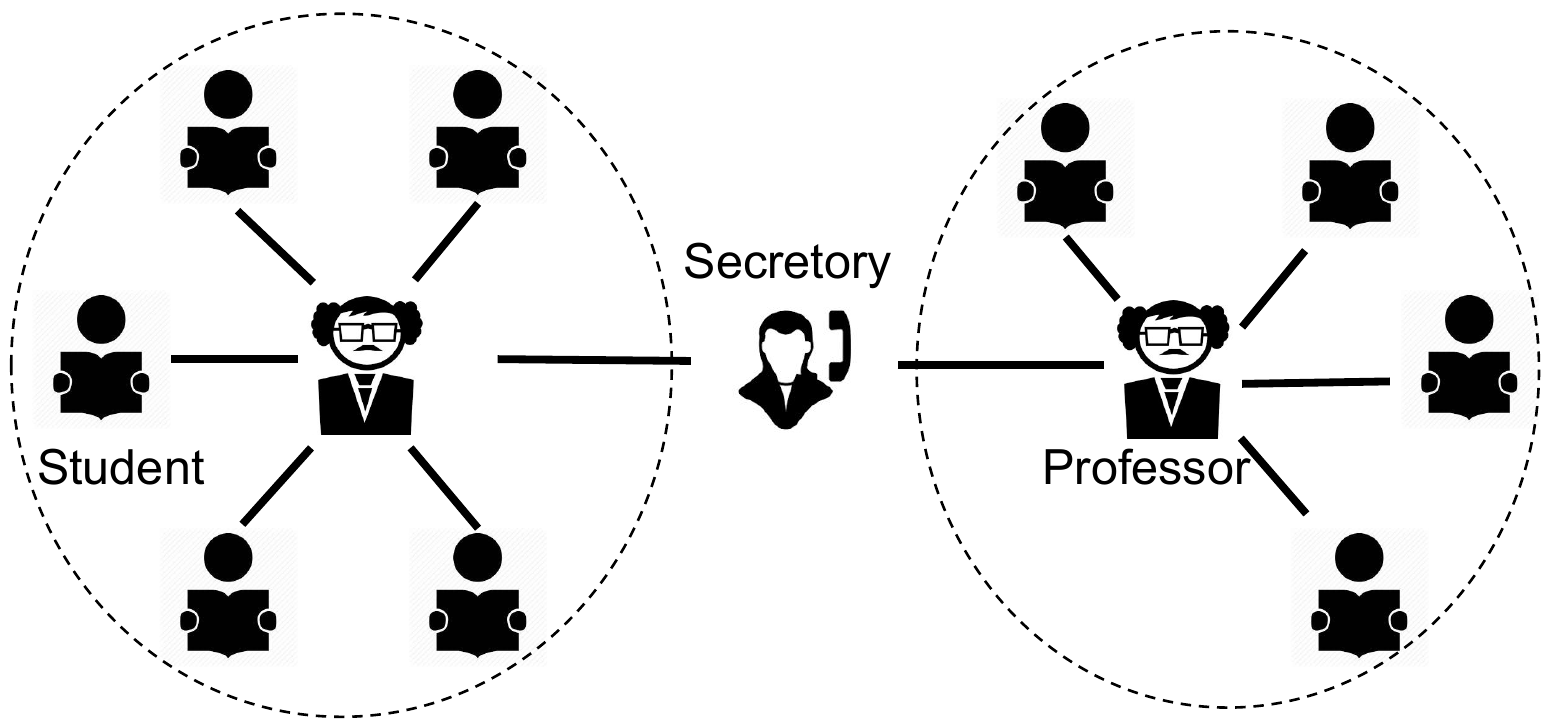}
\caption{\textbf{Toy example of communities in a CS department}}
\label{fig:toy}
\end{figure}

To see how these two assumptions do not necessarily hold true, consider the scenario in Figure \ref{fig:toy}, where two groups of students and professors are circled by dashed lines. Each of the two is naturally a valid community as everyone is affiliated to the same research group. However, they are not always densely connected, as it is possible that students work on their individual projects and hardly meet each other. The community is not homogeneous in a specific way either. Students may have similar age range, salary range and come from the same hometown, but these might be quite different from those of the professors. Therefore, unsupervised community detection algorithms based on edge density and node homogeneity can hardly detect communities like these.

Although neither of the density and homogeneity characteristics is prominent in the communities in Figure \ref{fig:toy}, we observe that there are indeed some interesting \textit{social patterns}. For example, a pattern of \textit{star} is clearly formed within the communities, and the secretory obviously acts as a \textit{bridge} between communities. Those patterns are characterized by both link structures and user contents. For instance, the professor as a star center is distinctive in ages from the students around but may possess similar other contents such as research interest and department. The professor is also densely connected with the students around, but the students may not directly work with each other. On the other hand, the secretory as a bridge shares similar contents like department with all others, while also connects to many of them. Nevertheless, she/he belongs to neither of the two groups. 
Clearly, communities are defined and should be detected by the underlying social patterns.

Intuitively, a social pattern as we want to leverage should be formed by users with certain contents into specific local structures, and it should appear often in communities.
However, while we can think of some intuitive ones like those in Figure \ref{fig:toy}, the actual social patterns should be \textit{fuzzy, complex and varying across networks}. It is impractical to enumerate all possible ones and design an unsupervised algorithm to detect communities of all patterns. 
Moreover, traditional ways of finding and matching patterns on even small networks are notoriously time-consuming and almost impossible on real-world large social networks \cite{fang2016semantic}. To the best of our knowledge, there exists no previous work that aims to leverage such fuzz, complex and varying patterns on networks.

In this work, we propose to inject weak supervision into community detection by learning an embedding model that automatically captures fuzzy social patterns. Instead of taking pre-defined assumptions about community characteristics, it is more reliable to learn from the data what the underlying social patterns look like. Specifically, given a network, we propose to leverage a few labeled communities as examples, and automatically explore and memorize their important social patterns. The model can then easily compute the embeddings on other parts of the network or other similar networks, which can then be leveraged for the detection of unlabeled communities there. For example, in Figure \ref{fig:toy}, it is intuitive to leverage the star pattern learned from one community to detect the other, and the patterns learned from research groups in the CS department can be well leveraged to detect research groups in other departments, shools, \etc.

While network embedding has attracted intense research attention recently \cite{grover2016node2vec, perozzi2014deepwalk, tang2015line,yang2015network,yang2016revisiting}, a supervised embedding directed by social patterns is non-trivial and has never been explored. The task is challenging due to the lack of an efficient way to explore and combine complex user contents, network structures and community labels into a unified learning framework that effectively captures social patterns. In this work, we develop an end-to-end deep architecture of Community Oriented Network Embedding (CONE). The key advantages of CONE over some existing network embedding techniques are as follows:

\begin{enumerate}[leftmargin=20pt]
\setlength\itemsep{0.05em}
\item \textit{Content expressive}: to deeply explore and understand high-dimensional noisy contents as signatures of social patterns, we start from content embedding, by decomposing it into multiple nonlinear layers of recurrent neural networks (RNN).
\item \textit{Structure aware}: to leverage links and capture social patterns with local structures, we design a network regularization layer, which essentially generalizes the embedding from single users to local structures according to high-order random-walk transitions on the graph.
\item \textit{Community oriented}: to exploit example communities, we connect a softmax supervision layer into an end-to-end framework to directly require users within the same communities and thus forming some special social patterns to be close in the embedded space. In this way, the supervision can be properly back propagated to the content embedding layer via the network regularization layer, which explores their mutual reinforcement and reliably captures prominent social patterns.
\item \textit{Out-of-sample}: CONE directly learns an embedding model, rather than the embeddings of a specific set of nodes. Therefore, it is able to handle the out-of-sample problem, which addresses the challenges of limited supervision and dynamic network.
\end{enumerate}


\section{Motivating Study}
\label{sec:analysis}
In this section, we show the deficiency of traditional unsupervised community detection algorithms by quantitively examining the density and homogeneity assumptions they adopt and their consequences.
On the contrary, we demonstrate that there are indeed some communities that are neither dense nor homogeneous, but they share some underlying social patterns. 

We conduct a set of analysis using the Facebook dataset described in \cite{mcauley2012learning}. This dataset includes 10 ego-networks, consisting of 193 social circles and 4,039 nodes. 10 ego users have manually identified all the communities to which their friends belong. The average number of social circles in each ego-network is 19, with an average community size of 22 users. We trust these manually labeled social circles and use them as the ground-truth communities. 

Let a graph $G=\{V, E\}$ represent a network with a vertex set $V$ and an edge set $E$.  A community can be represented as a sub-graph $C=\{V_C,E_C\}$ with $V_C \subset V$ and $E_C \subset E$. $\mathbf{a}_i$ is the content vector of a vertex $v_i \in V$. We evaluate the density and homogeneity of both ground-truth communities and communities detected by the state-of-the-art algorithms. 

{\flushleft {\bf Metrics.}}
We use a density metric $D(\cdot)$ and a homogeneity metric $H(\cdot)$ defined as follows:
\begin{align}
D(C) = \frac{2|E_C|}{|V_C|(|V_C| - 1)}.
\end{align}
The value of $D(C)$ ranges from 0 (a graph without edges) to 1 (a complete clique). It is a standard measure of the density of a network widely used in related literature \cite{yang2015defining}.
\begin{align}
H(C) = \frac{\sum\limits_{v_i,v_j \in V_C, v_i\neq v_j}\sigma(\frac{\mathbf{a}_i^T \mathbf{a}_j}{avg(\mathbf{a}_i^T\mathbf{a}_j)})}{|V_C|(|V_C|-1)},
\end{align}
where $avg(\mathbf{a}_i^T \mathbf{a}_j)$ is the average of content similarity among all pairs of nodes in the whole network $G$, and $\sigma(t)= \frac{1-e^{-t}}{1+e^{-t}}$ is the adjusted half sigmoid function in the range of $[0,1]$. The value of $H(C)$ also ranges from 0 (none of the nodes share the same contents) to 1 (all nodes share the same contents), while the value increases rapidly as the content similarity in $C$ is small compared with the average content similarity in the whole network $G$.

{\flushleft {\bf Compared algorithms.}} 
We study three mainstream community detection algorithms within the state-of-the-art: MinCut \cite{clauset2004finding} (based on network modularity), CESNA \cite{yang2013community} (based on probabilistic generative model) and InfoMaps \cite{rosvall2008maps} (based on information theory).

{\flushleft {\bf Density and homogeneity analysis.}}

\textit{Assumption 1}: nodes within the same community are densely connected.

\begin{figure}[h!]
\centering
\vspace{-10pt}
\subfigure[Ground-truth]{
\includegraphics[width=0.23\textwidth]{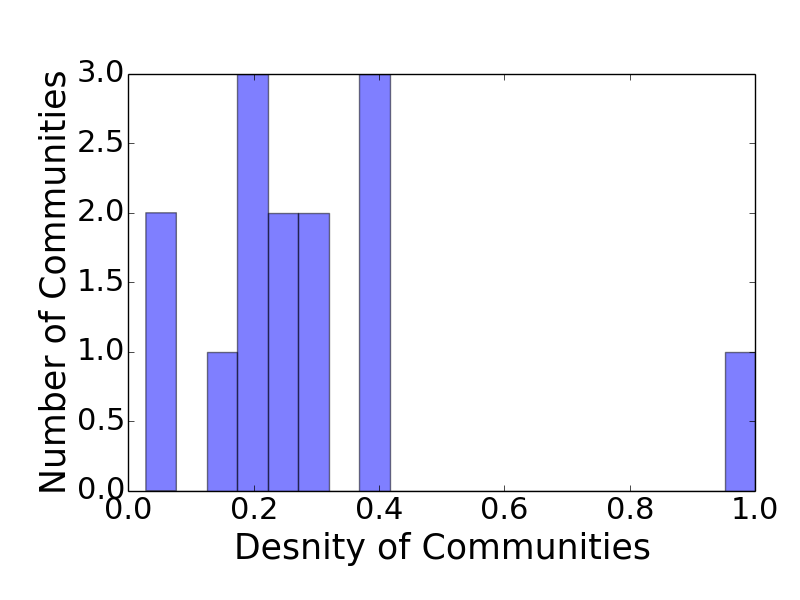}}
\subfigure[MinCut]{
\includegraphics[width=0.23\textwidth]{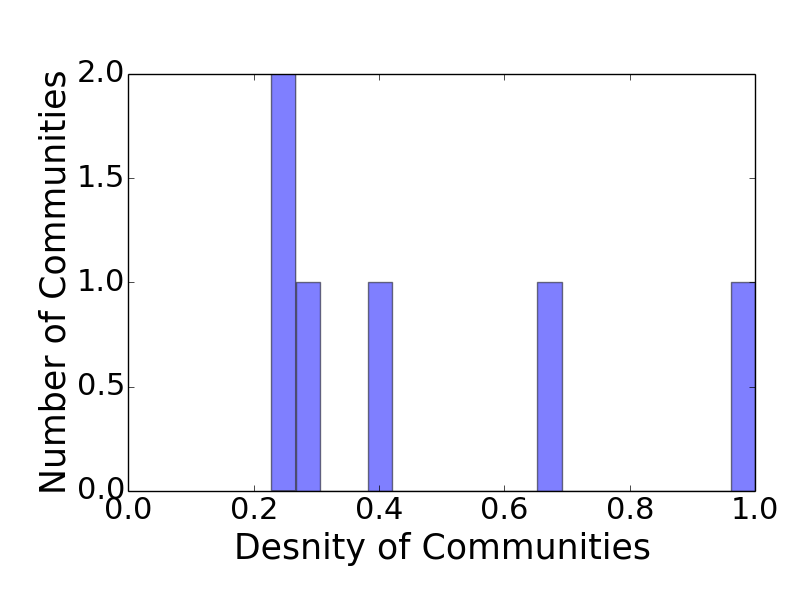}}
\subfigure[CESNA]{
\includegraphics[width=0.23\textwidth]{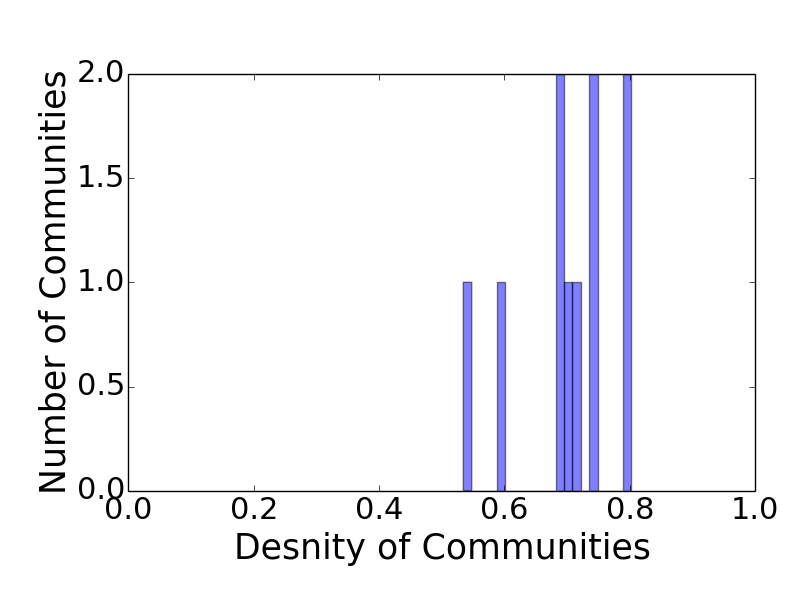}}
\subfigure[InfoMap]{
\includegraphics[width=0.23\textwidth]{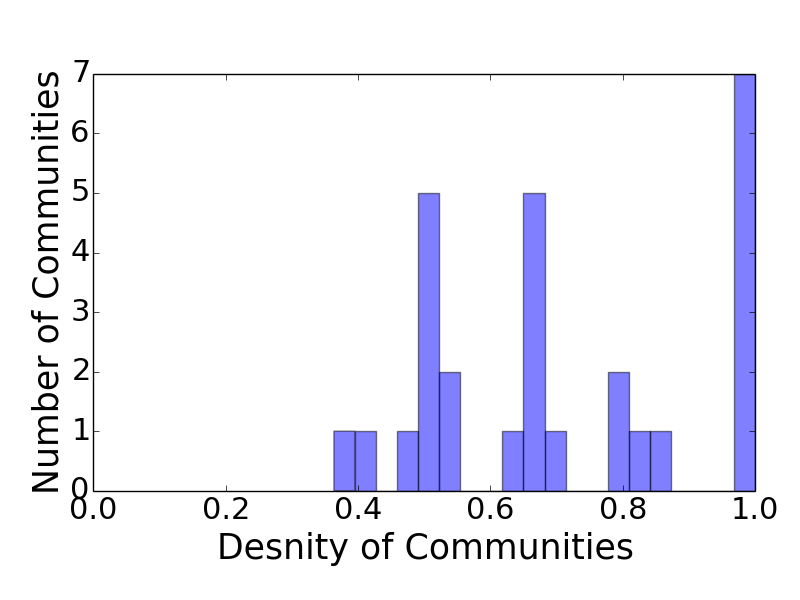}}
\caption{\textbf{Community density in ego-net 686.}}
\label{fig:dense}
\end{figure}

\textit{Conflicting evidence}:
In Figure \ref{fig:dense}, we present $D(C)$ computed on the ground-truth communities as well as the communities detected by the state-of-the-art algorithms in ego-network 686 as an example. It can be observed that ground-truth communities do not always have a high density. In fact, it is possible that there are more low-density communities than high-density ones as in (a). The evidence directly contradicts with Assumption 1.

Based on this inaccurate assumption, traditional unsupervised community detection algorithms tend to detect communities with high density.  As an example, the distributions of $D(C)$ computed on the detected communities in the same ego-network are presented in Figure \ref{fig:dense}(b)-(d). It can be observed that most of the detected communities have a high density ($D(C)>0.5$), which is inconsistent with the density distribution of the ground-truth communities. It suggests that the inaccurate assumption of edge density indeed leads to unreliable community detection results.

\begin{figure}[h!]
\centering
\vspace{-10pt}
\subfigure[Ground-truth]{
\includegraphics[width=0.23\textwidth]{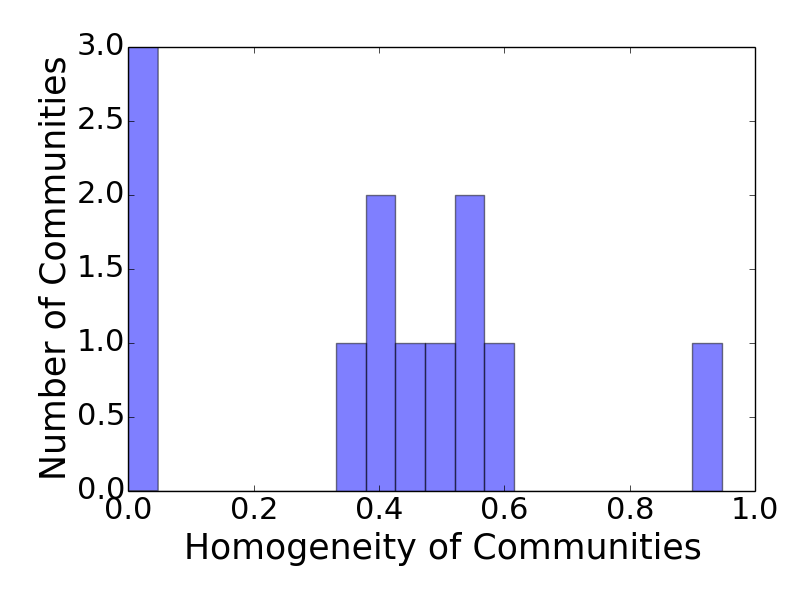}}
\subfigure[MinCut]{
\includegraphics[width=0.23\textwidth]{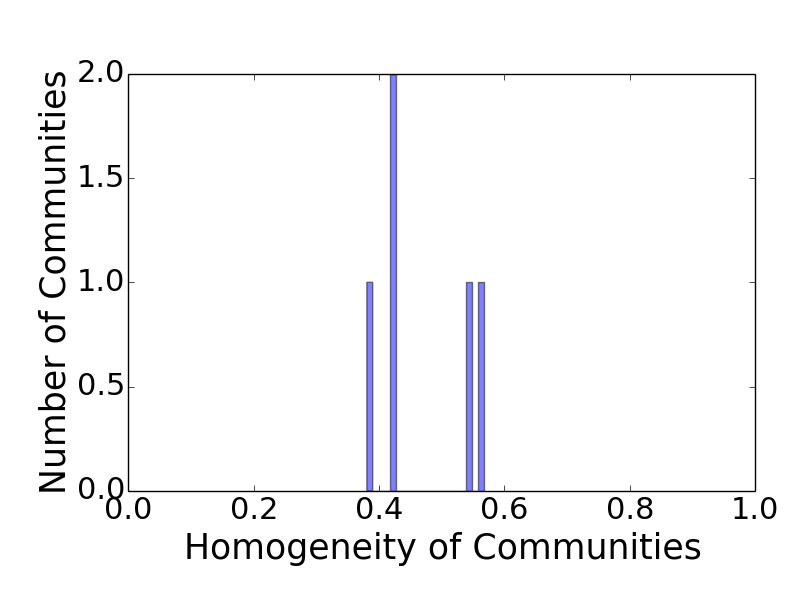}}
\subfigure[CESNA]{
\includegraphics[width=0.23\textwidth]{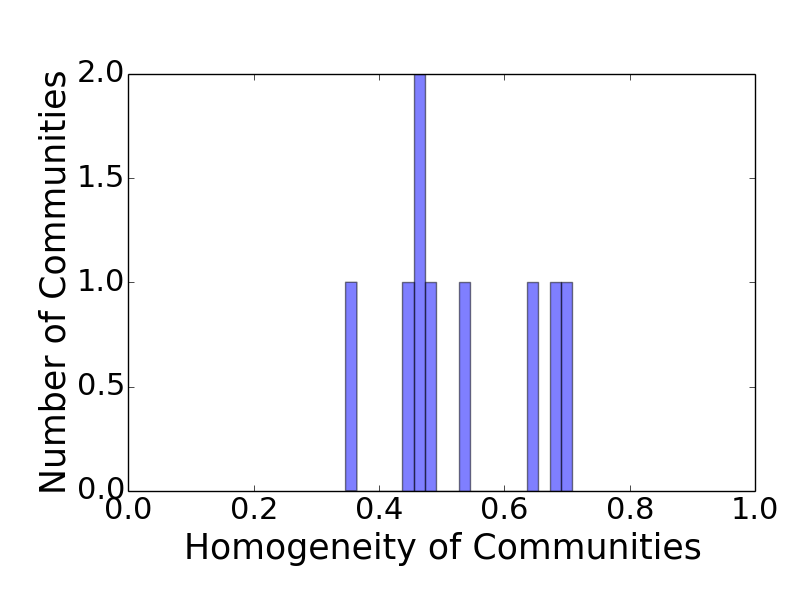}}
\subfigure[InfoMaps]{
\includegraphics[width=0.23\textwidth]{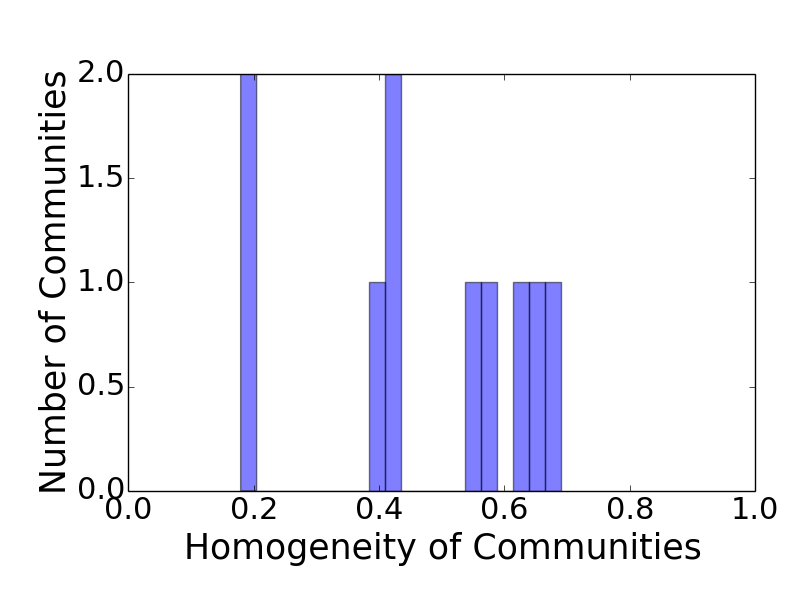}}
\caption{\textbf{Community homogeneity in ego-net 698.}}
\label{fig:homo}
\end{figure}

\textit{Assumption 2}: nodes within the same community are homogeneous in some ways.

\textit{Conflicting evidence}:
As an example, the distribution of $H(C)$ computed on the ground-truth communities in ego-network 698 is contradictory to Assumption 2. As shown in Figure \ref{fig:homo}(a), few communities are indeed homogeneous (\eg, the one with homogeneity higher than $0.8$), while the majority are much less homogeneous. Given $\sigma(1)\simeq 0.46$, the results indicate that most communities have a homogeneity similar to that of the whole network.

We conduct the same homogeneity analysis on communities detected by the same group of algorithms. As can be seen in Figure \ref{fig:homo}(c)-(d), since CESNA and InfoMaps assume node homogeneity in communities, they do detect communities of slightly higher homogeneity, which diverge from the ground-truth ones. MinCut does not assume node homogeneity.

\begin{figure}[h!]
\centering
\subfigure[Cmt 1 in ego-net 1684]{
\includegraphics[width=0.22\textwidth]{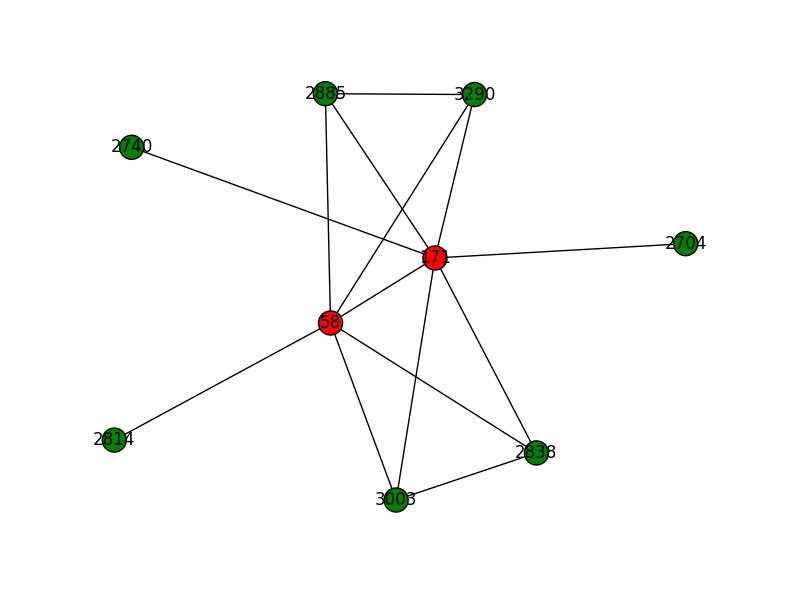}}
\subfigure[Cmt 15 in ego-net 3437]{
\includegraphics[width=0.22\textwidth]{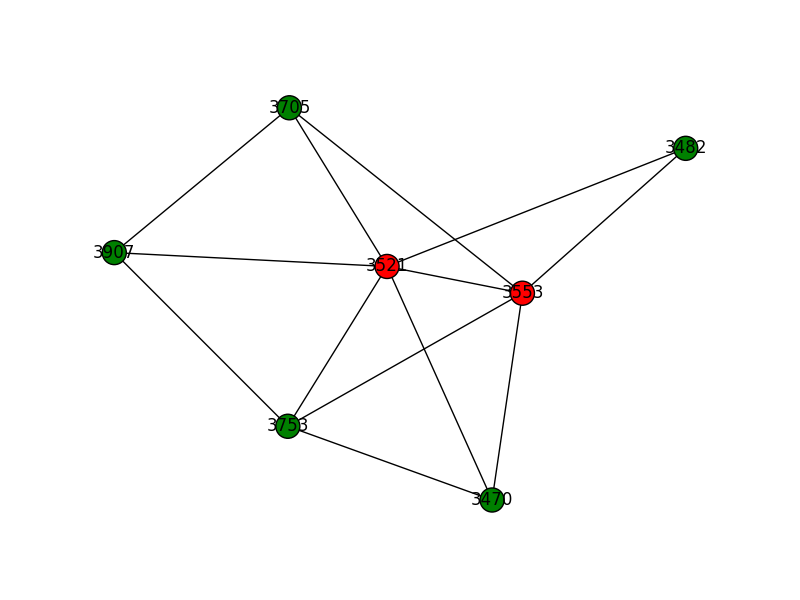}}
\caption{\textbf{Examples of real communities with social patterns.}}
\label{fig:pattern}
\end{figure}

{\flushleft {\bf Real community social pattern analysis.}}
After showing that nodes within the same communities are not always densely connected or homogeneous in some ways, we look into the structures of ground-truth communities to find evidence for the existence of social patterns, so as to further motivate our novel approach of community detection. 

In Figure \ref{fig:pattern}, among many patterns and examples, we show two communities that clearly share a \textit{co-star} pattern, where almost all members connect to two center nodes, while they do not densely connect within themselves. Moreover, the center nodes in both communities indeed have some special contents such as a certain degree in the college or job in the company, which are not shared by other non-center members. (We removed a few isolated noise nodes for clear visualization.)

\section{CONE}
\label{sec:model}
\subsection{Overall Framework}
The analysis in Sec.~\ref{sec:analysis} clearly demonstrates the deficiency of existing unsupervised community detection algorithms, which is a consequence of their falsifiable assumptions of edge density and node homogeneity. As we motivated in Sec.~\ref{sec:intro}, the existence of fuzzy, complex and varying social patterns underlying communities further indicates an urge for developing a novel community detection algorithm that aims at leveraging such patterns.

In this work, rather than enumerating and evaluating all important social patterns, we propose to automatically capture them from labeled communities through weak supervision. Instead of exhaustive graph matching, we model the problem as representation learning. Specifically, we do not care about the exact shapes of the patterns, but we intuitively require the users within the same patterns to be close in an embedded space. We formulate our framework as follows.

We are given a network modeled by $\mathcal{G}=\{\mathcal{V},\mathcal{E},\mathcal{A},\mathcal{C}\}$, where $\mathcal{V}$ is the set of $n$ users, $\mathcal{E}$ is the set of observed links among $\mathcal{V}$. $\mathcal{A}$ is the set of observed user contents on $\mathcal{V}$, where $\mathbf{a_i}$ is the set of contents on user $v_i$. $\mathcal{C}$ is a set of ground-truth community memberships, where $\mathbf{c}_k$ is the set of users in community $c_k$. For learning the model, only a small number of ground-truth communities are required as examples, while others are used for evaluation. 

To detect more communities, we firstly learn an embedding model that captures social patterns in $\mathcal{G}$ based on $\mathcal{A}$, $\mathcal{E}$ and $\mathcal{C}$. The model should effectively explore various social patterns and transform each user into a $p$-dimensional vector. In the embedded space, users forming some important social patterns and thus within each example community are close. More importantly, the model should be able to leverage the social patterns and produce embeddings on all users in $\mathcal{V}$ in a similar way. Therefore, new communities can then be detected through generic clustering algorithms such as $k$-means on the $p$-dimensional features. Overlapping communities can also be discovered by algorithms like MOC \cite{banerjee2005model}.

\begin{figure*}[h!]
\centering
        \includegraphics[width=1\linewidth]{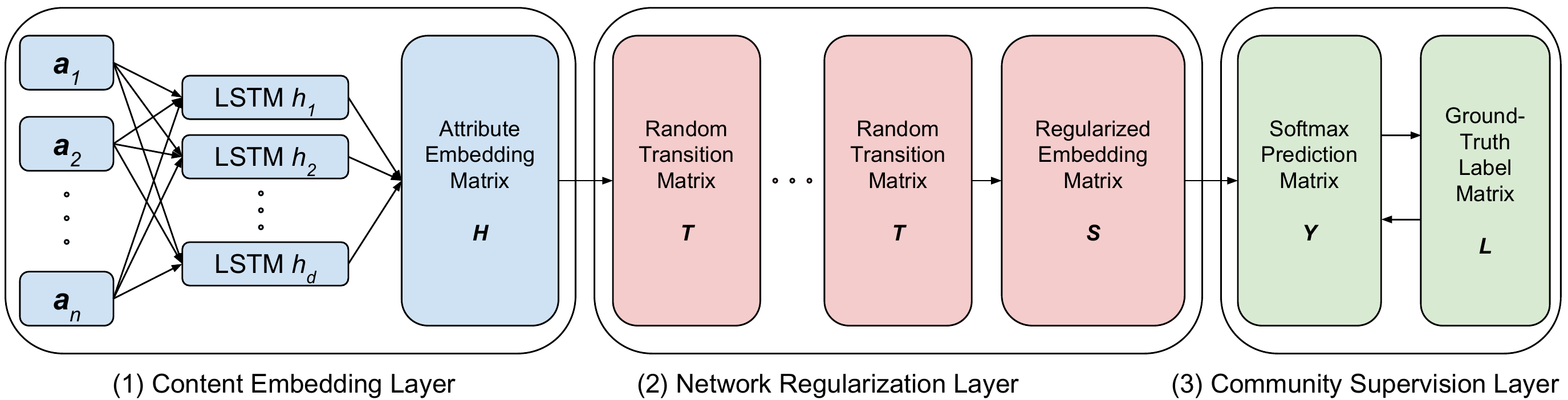}
    \caption{The end-to-end deep architecture of Community Oriented Network Embedding (CONE).}
    \label{fig:arch}
 \end{figure*}
 
Figure \ref{fig:arch} illustrates the overall architecture of our CONE model. 
We take the input of $\mathcal{A}$ for all users $\mathcal{V}$ and compute the content embeddings $\mathbf{H}$, which is then regularized by network structures $\mathcal{E}$ to yield the community oriented embeddings $\mathbf{S}$. For representation learning, $\mathbf{S}$ is then used to generate community predictions $\mathbf{Y}$, which is supervised by the ground-truth community labels $\mathbf{L}$ derived from the example communities in $\mathcal{C}$. $\mathbf{L}$ can be either point-wise, where $l_{ik}=1$ means $v_i$ is a member of community $c_k$, or pair-wise, where $l_{ij}=1$ means $v_i$ and $v_j$ are in the same community. To detect more communities, we take $\mathbf{S}$ as the actual output of CONE and apply generic clustering algorithms like $k$-means on it.

To ensure a desirable representation, ideally we should require users within the same example communities indicated by $\mathbf{L}$, and thus forming some specific social patterns, to be close in the embedded space. That is, we need to compute the pair-wise loss among all users and our overall high-level objective function should look like the following, 
\begin{align}
\mathcal{J}=\Phi(\mathbf{Y},\mathbf{L})=\sum_{(v_i,v_j), i\neq j}\phi(\hat{y}_{ij}, l_{ij}).
\label{eq:highobj}
\end{align}
where $\Phi$ is a loss function, $\mathbf{L}$ is the pair-wise community labels and $\mathbf{Y}$ is the pair-wise prediction.

In what follows, we further explain the reasons for our model architecture and how it works in details.

\subsection{Deep Architecture}
While it is intuitive to automatically learn an embedding model that captures important social patterns, the task is non-trivial and posts some unique challenges:
\begin{itemize}[leftmargin=20pt]
\setlength\itemsep{0.05em}
\item Explore high-dimensional noisy user contents.
\item Leverage complex links and local structures.
\item Exploit limited supervision of example communities.
\end{itemize}
To deal with all challenges above, we develop a deep architecture of CONE, which inherently combines RNN with random-walks in an end-to-end supervised learning framework.

{\flushleft {\bf RNN: deeply explore and understand user contents as signatures of social patterns.}}
Existing network embedding algorithms are insufficient in exploring contents for capturing social patterns. They usually focus on preserving the link structures among users \cite{grover2016node2vec, tang2015line, perozzi2014deepwalk}, and incorporate user contents as augmented attribute nodes \cite{tang2015pte}, text feature matrices \cite{yang2015network} or bag-of-word vectors \cite{yang2016revisiting}. Thus, the deep semantics within user contents are not fully explored. 
In our situation, contents are so important that they often become the signatures of social patterns. For example, in a football fans' club, a popular player identified by the contents of his tweets is likely to be the center, surrounded by groups of fans identified by their semantically different tweet contents. 

However, exploring and understanding user contents in social networks is a non-trivial task, since they can be complex, noisy and high-dimensional.
\xeg, in text-rich networks like Twitter, contents can be a list of recent tweets; in tag-rich networks like Flickr, they can be a list of frequently used hashtags in a timely order; in networks with explicit attributes like Facebook and LinkedIn, they can be a set of categorical variables like \textit{Birthday, School, Occupation} with lots of noisy and missing data.
The situation reminds us of the task in natural language processing (NLP) of understanding noisy text sequences, where the semantics is usually hidden in ordered tokens of variable lengths.

In this work, instead of starting from the links, we start from deeply modeling user contents. To this end, we employ RNN from NLP, which has been proven advantageous over various other methods in understanding text-like sequences, due to its supreme expressiveness within the neural networks to explore, understand and memorize important semantic patterns.
The deep learning framework of RNN also provides us with the flexibility to modify the neural network architectures in order to leverage other information like network structures and community supervision. 
To the best of our knowledge, this is the first work that models user contents as raw texts and successfully applies RNN to network embedding. 

To leverage RNN, given a user $v_i$'s textual contents such as a list of recent tweets, we can concatenate them into a single sequence $\mathbf{s}_i$, each element of which is the index of the corresponding word after stemming and stop word removing; given $v_i$'s categorical contents  $\mathbf{a}_i$ such as \textit{education} and \textit{work} on Facebook, we transform it into a sequence $\mathbf{s}_i=\{j| a_{ij} = 1\}$. The sequence $\mathbf{s}_i$ is then used as the input of RNN. As mentioned in \cite{sutskever2014sequence}, simple RNN would be difficult to train due to the resulting long term dependencies. Therefore, we use the long-short term memory (LSTM) cells \cite{hochreiter1997long} instead to embed $\mathcal{A}$. The architecture and implementation of LSTM can be found in the public website\footnote{http://deeplearning.net/tutorial/lstm.html}. 

Upon each input $\mathbf{s}_i$, there will be one output from the LSTM cell as a semantic embedding of user contents. To further improve model expressiveness, we use $d$ LSTM cells to output $d$ embedding vectors $\{\mathbf{h}_i^1, \mathbf{h}_i^2,\ldots, \mathbf{h}_i^d\}$, and apply mean pooling on them, which is commonly used to integrate features and reduce deviation. So we have
\begin{align}
\mathbf{h}_i=h(\mathbf{s}_i) = \frac{1}{d}\sum_{j=1}^d \mathbf{h}_i^j,
\label{eq:h}
\end{align}
where $h(\cdot)$ denotes the overall deep content embedding function. Supposing the embedding size of each LSTM cell is $p$, we get a content embedding matrix $\mathbf{H} \in \mathbb{R}^{p\times n}$ to represent the embeddings of $\mathcal{V}$, where the $i$-th column of $\mathbf{H}$ equals to $\mathbf{h}_i$.

Note that, while RNN is especially useful for exploring complex text-like sequences, for simple numerical or categorical contents like user attributes, it also makes sense to use simpler models like feedforward neural networks, which can be incorporated into our end-to-end framework in the same way. We will also show the performance of such basic neural networks in our experiments.

{\flushleft {\bf Random-walk: leverage links and capture social patterns with local structures.}}
The objectives of existing network embedding algorithms are not appropriate for leveraging network structures to capture social patterns. 
They usually model network structures by sampling a set of paths from the networks and applying a Skipgram-based model \cite{mikolov2013distributed} to uniformly require the embeddings of nodes that share similar graph context to be similar \cite{grover2016node2vec,perozzi2014deepwalk,  yang2015network, yang2016revisiting}.
To capture social patterns, we want the user embedding to be guided by community supervision. Users having similar network context do not necessarily belong to the same communities, and thus should not be required to have similar embeddings without discrimination. 
Besides, by sampling the networks into paths, structural information beyond paths is not efficiently leveraged. In our situation, the shapes of local network structures are extremely important, since they may well indicate the existence of specific social patterns, like the star and co-star shapes as we mentioned in Sec.~\ref{sec:intro} and \ref{sec:analysis}.

In this work, we directly leverage local structures to regularize the supervision of example communities. To this end, we armor our RNN-based content embedding with random-walk-based network regularization. It efficiently generalizes the embeddings on single users to their ambient local structures, allowing RNN to explore content embedding under the regularization of network local structures and the guidance of example communities.

Consider the RNN-based content embedding. Due to Eq.~\ref{eq:h}, if $\mathbf{a}_i$ is similar to $\mathbf{a}_j$, then it is likely that the embeddings $\mathbf{h}_i$ and $\mathbf{h}_j$ are also close.
However, this may not be ideal, because similar users can well form different communities. 
The key question is, are they also on the same local structure? 

To account for this, we insert a random-walk-based network regularization layer, which recomputes the embedding of each user \wrt~her neighbors, according to their distances measured by random-walks of $k$ steps, \ie, $\mathbf{s}_i=\sum_{j=1}^n t^k_{ji}\mathbf{h}_j$, where $t^k_{ji}$ is the $k$-step random-walk transition probability from $v_j$ to $v_i$. In matrix form, a more compact formula is $\mathbf{S}=\mathbf{H}\mathbf{T}^k$. $t_{ij}=w_{ij}/d_i$, where $w_{ij}$ is the binary or real-valued weights on edge $e_{ij}$ and $d_i=\sum_{j}w_{ij}$.

Intuitively, each $v_i$ `collects' the embeddings transmitted from its local neighbors. 
Compared with $\mathbf{h}_i$, $\mathbf{s}_i$ encodes the local structure around $v_i$, rather than just the semantic information from $\mathbf{a}_i$. As a consequence, only users with similar contents as well as local structures will be embedded as close.

Moreover, consider the loss in Eq.~\ref{eq:highobj} brought by the supervision of example communities. Without network regularization, the loss on content embeddings can be formulated as
\begin{align}
\mathcal{J}_H=\sum_{(v_i,v_j),i\neq j} \phi(\hat{y}(h(\mathbf{a}_i),h(\mathbf{a}_j)), l_{ij}).
\end{align}
Under $\mathcal{J}_H$, \eg, if there are some professors and students in the same example communities, all professors and students will be required to get similar content embeddings, which should not be the case.
Instead, we apply supervision on the regularized embeddings as
\begin{align}
\mathcal{J}_S=\sum_{(v_i,v_j),i\neq j} \phi(\hat{y}(\mathbf{s}_i,\mathbf{s}_j), l_{ij}).
\end{align}
As a consequence, only professors and students connected in certain local structures indicated by supervision are embedded as close.

{\flushleft {\bf End-to-end learning: exploit community supervision and the mutual reinforcement between contents and links.}}
As we discussed about Eq.~\ref{eq:highobj} before, to directly meet our goal of embedding users in the same example communities to be close, we need to construct a pair-wise loss on each pair of users in the same communities. However, pair-wise loss functions can not be built into our end-to-end embedding framework and permit efficient training of the neural networks. 
Moreover, converting the community memberships to pair-wise 0-1 labels will lead to significant information loss, especially for overlapping communities. 

To overcome these difficulties, we find the point-wise softmax prediction with cross entropy loss as a suitable substitute to the exact pair-wise loss \cite{bishop2006pattern}. 
While having a different objective, softmax with cross entropy basically ensures that instances predicted with the same label are close in a space that can be viewed as a linear projection of the original embedding space. Therefore, it can be viewed as achieving a similar goal as the direct pair-wise loss. 
Moreover, softmax is commonly used in end-to-end deep learning frameworks to predict multiple labels, and the cross entropy loss can be efficiently back propagated through neural networks. 
By allowing the training on multi-label predictions, it is also able to leverage overlapping example communities and distinguish among different communities. 

Therefore, we incorporate a softmax supervision layer to form our end-to-end deep learning framework,  where community supervision can be exploited as a guidance for exploring social patterns, and the mutual reinforcement between user contents and link structures can be leveraged.
Our practical objective function based on cross entropy loss is as follows: 
\begin{align}
\hat{y}_{ik} = \frac{\exp (\mathbf{w}^T_k\mathbf{s}_i)}{\sum_{k'=1}^K\exp(\mathbf{w}^T_{k'}\mathbf{s}_i)},\quad
\mathcal{J} = \sum_{i=1}^n\sum_{k=1}^K l_{ik} log(\hat{y}_{ik}),
\label{obj}
\end{align}
where $\mathbf{w}$'s are the model parameters in the softmax layer and $K$ is the total number of example communities. $\mathbf{Y}$ and $\mathbf{L}$ encode the predicted and ground-truth community labels, respectively, where $\hat{y}_{ik}=1$ means $v_i$ is predicted to be a member of community $c_k$, and $\mathbf{L}$ is the point-wise community labels.

To optimize Eq.~\ref{obj}, we employ stochastic gradient descent (SGD) with the diagonal variant of AdaGrad from \cite{qiu2015convolutional}. At the $t$-th step, the parameters $\Theta$ is updated by:
\begin{eqnarray}
\Theta_{t} \gets \Theta_{t-1}-\frac{\rho}{\sqrt{\sum_{i=1}^{t}g_{i}^{2}}}g_{t},
\end{eqnarray}
where $\rho$ is the initial learning rate and $g_{t}$ is the sub-gradient at time $t$. Since our network regularization layer is implemented as a matrix multiplication, the gradients can be efficiently back-propagated to the content embedding layer, and the overall embedding framework is end-to-end. 

When detecting communities in a network, embeddings of users produced by the CONE model learned on labeled data are fed into generic clustering algorithms like $k$-means. We apply cross-validation to automatically choose the optimal number of communities as done in \cite{yang2013community}. 

{\flushleft {\bf The efficiency of CONE:}} 
CONE is an end-to-end deep learning framework implemented using TensorFlow\footnote{https://www.tensorflow.org/}. Only minimum setup is required to run it efficiently on GPU. 
We will make the code available upon acceptance of the work.
\section {Experiments}
\label{sec:exp}
In this section, we evaluate CONE for community detection with extensive experiments on 3 real-world networks.

\subsection{Experimental Settings}
{\flushleft \bf Datasets.}
We use three real-world network datasets. 
The first is the Facebook dataset we used in Sec.~\ref{sec:analysis} for data analysis, which consists of 10 ego-networks with community labels explicitly provided by the ego users \cite{mcauley2012learning}. The contents in this dataset are well-defined user profiles such as \textit{education, work} and \textit{location}, and links are undirected friendships among users. 
The second is a Flickr dataset collected by \cite{Wang-etal12}. The community labels are generated from the groups joined by users. The contents are the tags aggregated on users' posted photos and the links are undirected friendships.
The third is a Twitter dataset also collected by \cite{mcauley2012learning}, which consists of 973 ego-networks. The community labels are generated from friend circles (or lists), \ie, friends put into the same list by the ego user are regarded as within one community. The contents are the hashtags and popular account mentions within users' recent tweets and the links are directed followings. 
Detailed statistics of the three datasets we use are shown in Table 1.


\begin{table}[h!]
\centering
 \begin{tabular}{|c|cccc|}
   \hline
Dataset&\#nodes&\#comms&\#links&\#attrs\\
  \hline
Facebook& 4,039 & 192 & 88,234 & 634\\
\hline
Twitter& 81,306 & 837 & 1,768,149 & 12,274\\
\hline
Flickr& 35,313 & 200 & 3,017,530 & 77,263\\
\hline
 \end{tabular}
 \label{tab:stat}
 \caption{\label{T2}\textbf{Statistics of 3 real-world network datasets.}}
\end{table}

{\flushleft \bf Compared algorithms.} 
We compare with two groups of algorithms from the state-of-the-art to comprehensively evaluate the performance of CONE.
{\flushleft \it Community detection algorithms.} Some algorithms are based on the edge density assumption alone, while others also assume node homogeneity. We compare with this group of algorithms to show the advantage of abandoning these inaccurate assumptions and leveraging the supervision of example communities.
\begin{itemize}[leftmargin=20pt]
\setlength\itemsep{0.1em}
\item \textbf{MinCut} \cite{clauset2004finding}: a classic community detection algorithm based on modularity.
\item \textbf{BigClam} \cite{yang2013overlapping}: an advanced community detection algorithm solely based on network structure.
\item \textbf{Circles} \cite{mcauley2012learning}: a generative model of edges \wrt~attribute similarity to detect communities.
\item \textbf{CESNA} \cite{yang2013community}: a generative model of edges and attributes to detect communities.
\item \textbf{SGM} \cite{eaton2012spin}: a semi-supervised framework incorporating individual labels and pair-wise constraints.
\end{itemize}
{\flushleft \it Network embedding algorithms.} While we find it intuitive to model the problem as representation learning, we compare with the state-of-the-art network embedding algorithms to show that CONE is advantageous in capturing social patterns.
The embeddings learned by all algorithms are fed into the same $k$-means clustering algorithm as CONE to produce community detection results.
\begin{itemize}[leftmargin=15pt]
\setlength\itemsep{0.1em}
\item \textbf{DeepWalk} \cite{perozzi2014deepwalk}: an embedding algorithm based on truncated random walks that only considers network structures.
\item \textbf{node2vec} \cite{grover2016node2vec}: an embedding algorithm based on $2nd$ order random walks that only considers network structures.
\item \textbf{TADW} \cite{yang2015network}: an embedding algorithm that generalizes DeepWalk to consider both node attributes and network structures by tensor factorization.
\item \textbf{PTE} \cite{tang2015pte}: an embedding algorithm that generalizes LINE \cite{tang2015line} to consider node attributes, network structures and class labels by graph augmentation.
\item \textbf{Planetoid} \cite{yang2016revisiting}: an embedding algorithm that extends DeepWalk to consider node features, network structures and class labels by jointly predicting labels and contexts.
\end{itemize}
The number of communities to detect is tuned via standard 5-fold cross validation for all algorithms. 
The implementations of compared algorithms are all provided by the original authors.
{\flushleft \bf Metrics.}
Two widely used metrics for evaluating community detection results are used in our experiments.
For a detected community $c_i^*$ and a ground-truth community $c_i$, the \textit{F1 similarity} and \textit{Jaccard similarity} are defined as
\[F1 = \frac{2 \cdot precision \cdot recall}{precision + recall},\]
\[Jaccard = \frac{|c_i \cap c^*_i|}{|c_i \cup c^*_i|},\]
where $precision = \frac{c_i \cap c^*_i}{|c^*_i|}$, $recall = \frac{c_i \cap c^*_i}{|c_i|}$. 

For a set of ground-truth communities $\{c^*_i\}_{i=1}^M$ and a set of detected communities $\{c_i\}_{i=1}^N$, we compute the score as
\[\frac{1}{2|C|} \cdot \sum_{c_i \in C} \max_{c^*_i \in C^*} eval(c_i, c^*_i) + \frac{1}{2|C^*|} \cdot \sum_{c^*_i \in C^*} \max_{c_i \in C} eval(c_i, c^*_i),\]
where $eval(c_i, c^*_i)$ can be either replaced by $F1$ or $Jaccard$.

\subsection{Performance Comparison with Baselines}
We quantitively evaluate CONE against all baselines on community detection. We randomly split the labeled communities into training and testing sets. 
We use $10\%$ of the labeled communities as examples to learn the models for CONE, SGM, PTE and Planetoid.
All compared algorithms are then evaluated on the rest $90\%$ labeled communities. 
To observe significant difference in performance, we split the training and testing sets 10 times and conduct statistical t-tests with $p$-value 0.01. 

\begin{table}[h]
\small
 \begin{tabular}{|c|ccc|ccc|}
 \hline
\multirow{2}{*}{Algorithm}&\multicolumn{3}{c|}{F1 Score}&\multicolumn{3}{c|}{Jaccard Score}\\
\cline{2-7}
&FB&Flickr&Twitter&FB&Flickr&Twitter\\
\hline
MinCut&0.265&0.059&0.114&0.181&0.031&0.081\\
\hline
BigClam&0.284&0.076&0.121&0.198&0.041&0.084\\
\hline
Circles&0.267&0.042&0.102&0.185&0.029&0.069\\
\hline
CESNA&\textit{0.298}&0.076&0.126&\textit{0.242}&0.042&0.084\\
\hline
SGM&0.281&0.077&0.177&0.202&0.043&0.116\\
\hline
DeepWalk&0.211&0.071&0.136&0.135&0.039&0.094\\
\hline
node2vec&0.245&0.074&0.119&0.145&0.041&0.084\\
\hline
TADW&0.269&0.073&0.146&0.193&0.036&0.105\\
\hline
PTE&0.221&0.069&0.147&0.142&0.038&0.110\\
\hline
Planetoid&0.272&\textit{0.079}&\textit{0.188}&0.231&\textit{0.045}&\textit{0.121}\\
\hline
\textbf{CONE}&\textbf{0.397}&\textbf{0.096}&\textbf{0.206}&\textbf{0.313}&\textbf{0.052}&\textbf{0.149}\\
\hline
 \end{tabular}
 \caption{\label{tab:perform}\textbf{Performance comparison on 3 datasets.}}
\end{table}

Table \ref{tab:perform} shows the average F1 and Jaccard scores evaluated on all compared algorithms over the same 10 random splits of training and testing sets. The results all passed our t-tests. The parameters of baselines are all set to the default values as suggested in the original works. Some numbers are a bit different with those in the original work, because we directly use all node contents as input and only evaluate on the testing set. For CONE, we intuitively use two random transitions in the regularization layer to stress network locality and leverage the link structures in close neighborhoods. 

CONE reaches around 30\% relative improvements on the Facebook dataset and more than 10\% improvements on the other two datasets, compared with the second-runner among the 10 baselines on both F1 and Jaccard scores, while they have varying performance. This indicates the robustness and general advantages of our proposed approach. 

Taking a closer look, we observe that the scores on the Facebook dataset look much better than those on the other two. This is due to the high quality of both contents and links in the dataset. Since the Facebook dataset is the only one that has explicit human labels of communities, the large performance improvement on it indicates the advantage of CONE in leveraging user provided community examples to understand specific social patterns. 

The scores of network embedding algorithms are lower than traditional community detection algorithms on the Facebook dataset, probably because traditional models work better on few and clean contents. However, as the contents become high-dimensional and noisy like in the Flickr and Twitter datasets, network embedding algorithms excel. This indicates the advantage of leveraging embeddings for traditional social network tasks and further proves the efficacy of CONE in dealing with complex user contents.

All experiments are done on a local PC with two 2.5 GHz Intel i7 processors and 8GB memory. The runtime of CONE is comparable to all baselines, while it is trivial to run more efficiently on GPU.
\subsection{Model Selection}
We comprehensively evaluate the performance of CONE with different amounts of supervision and varying neural architecture.

\begin{figure*}[h!]
\centering
\subfigure[Facebook]{
\includegraphics[width=0.335\textwidth]{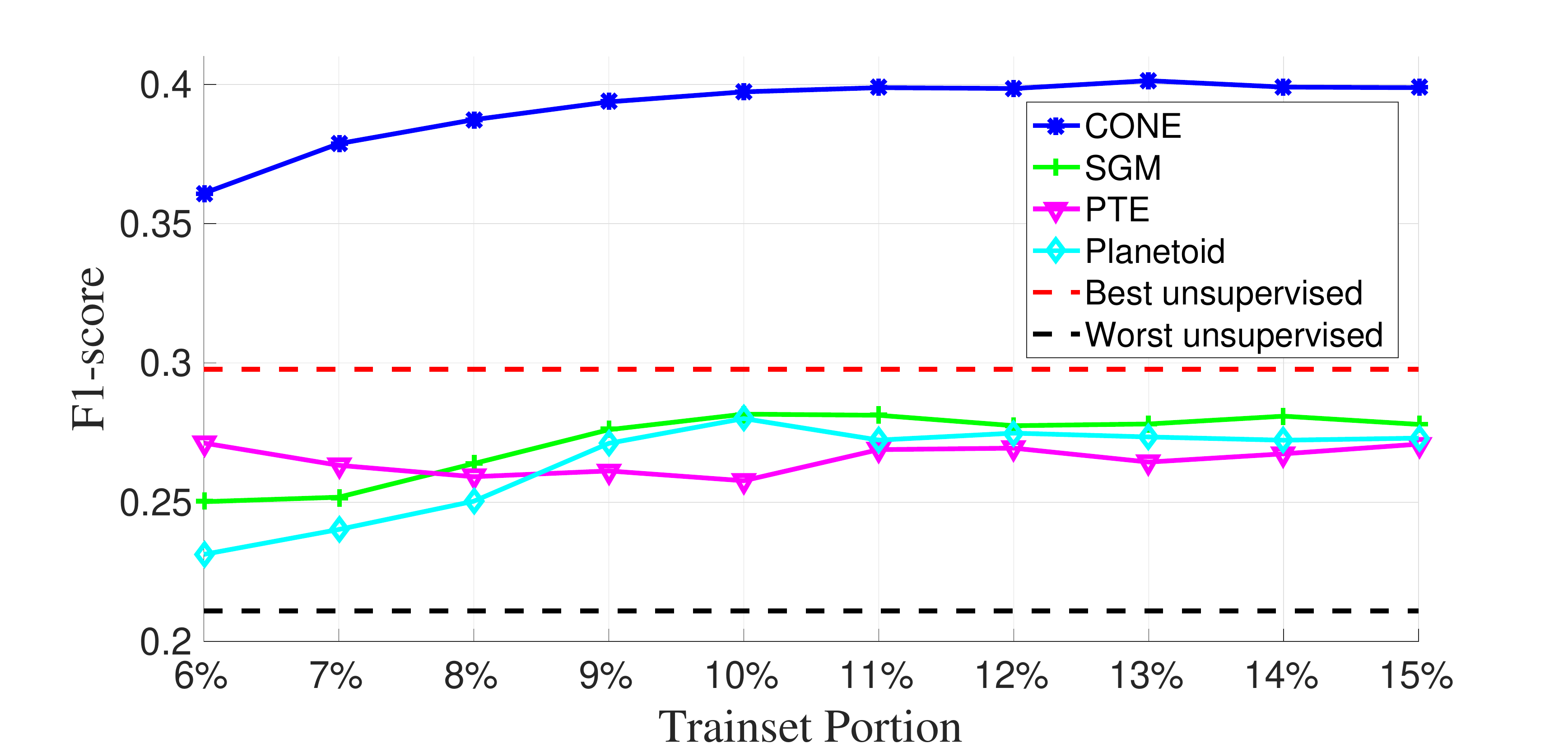}}
\hspace{-15pt}
\subfigure[Flickr]{
\includegraphics[width=0.335\textwidth]{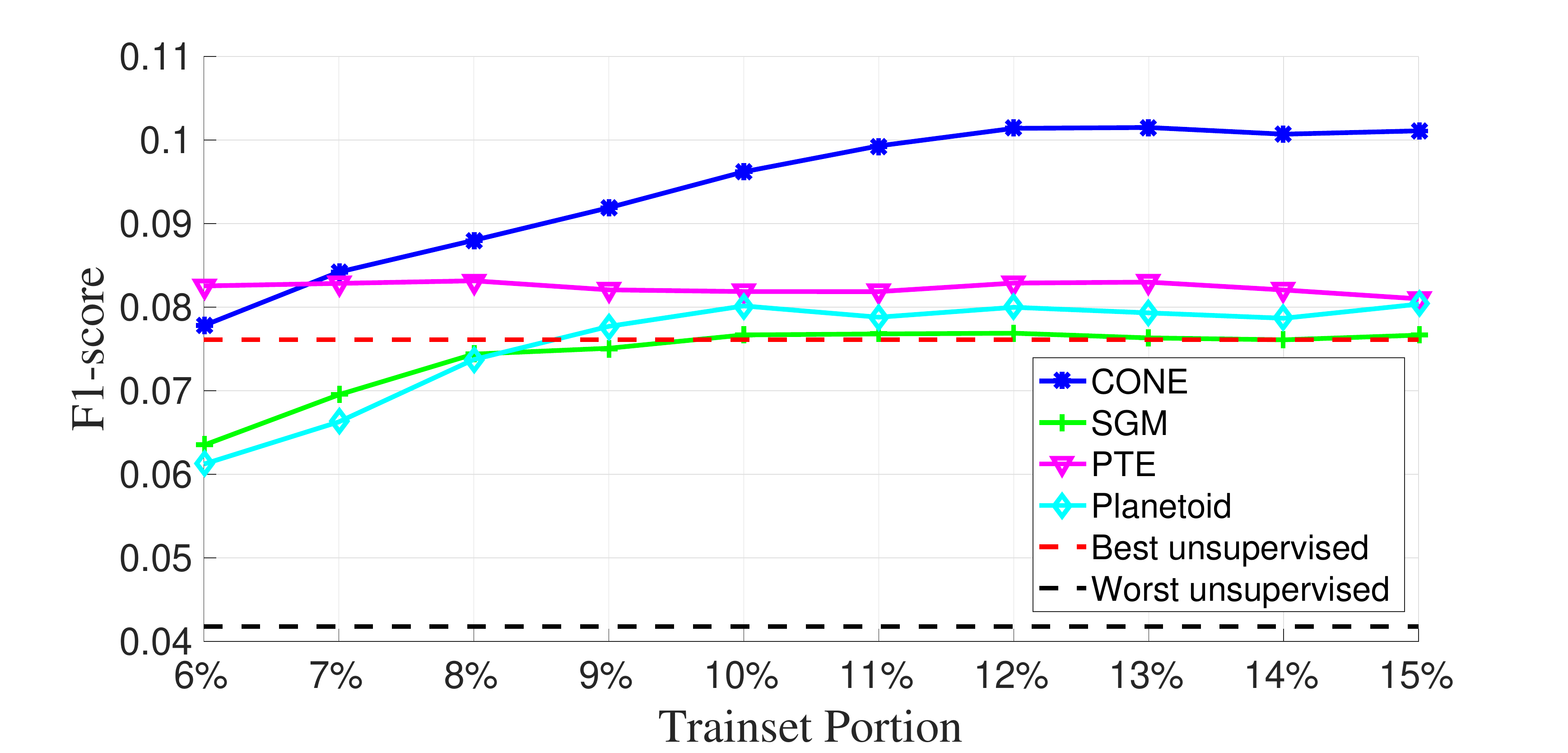}}
\hspace{-15pt}
\subfigure[Twitter]{
\includegraphics[width=0.335\textwidth]{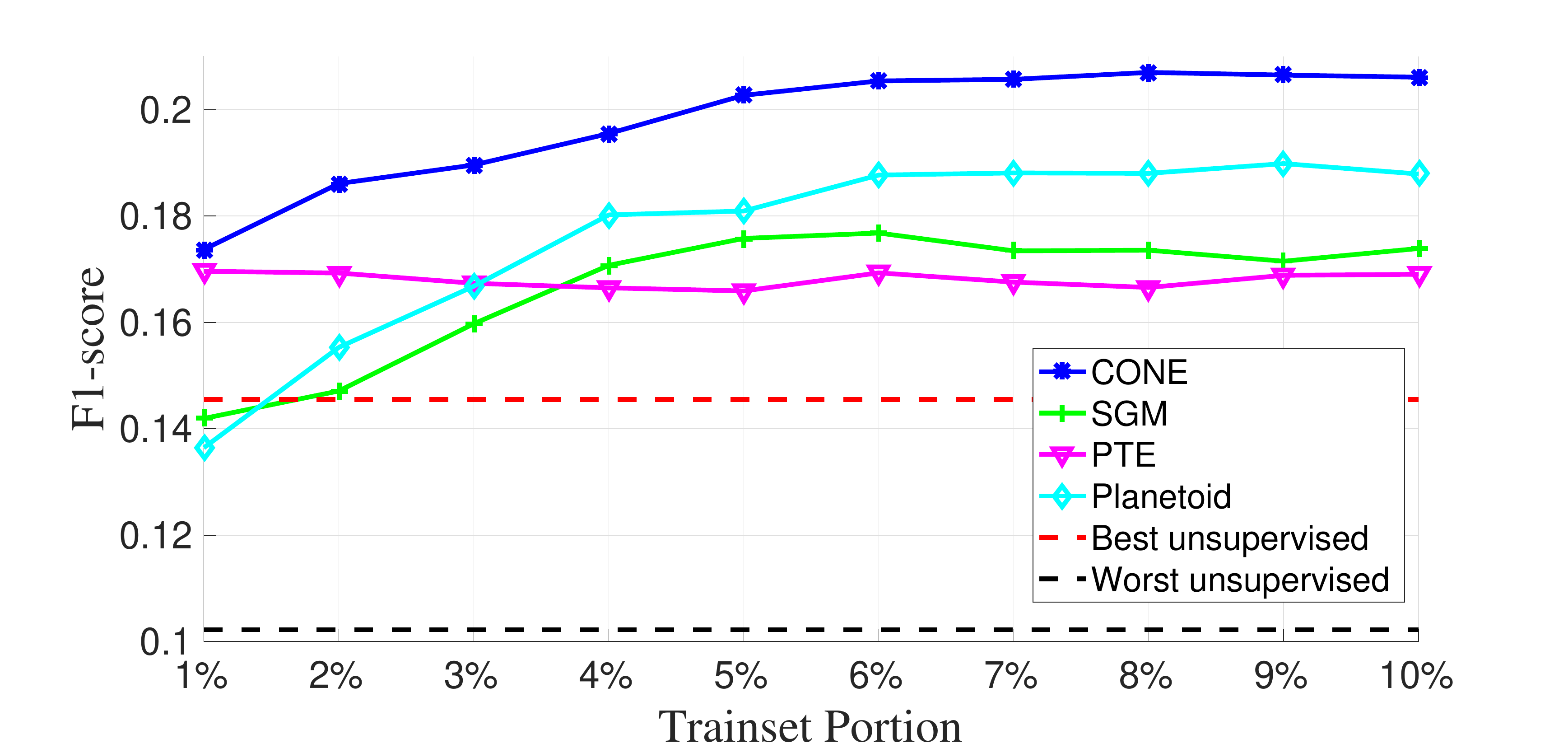}}
\subfigure[Facebook]{
\includegraphics[width=0.335\textwidth]{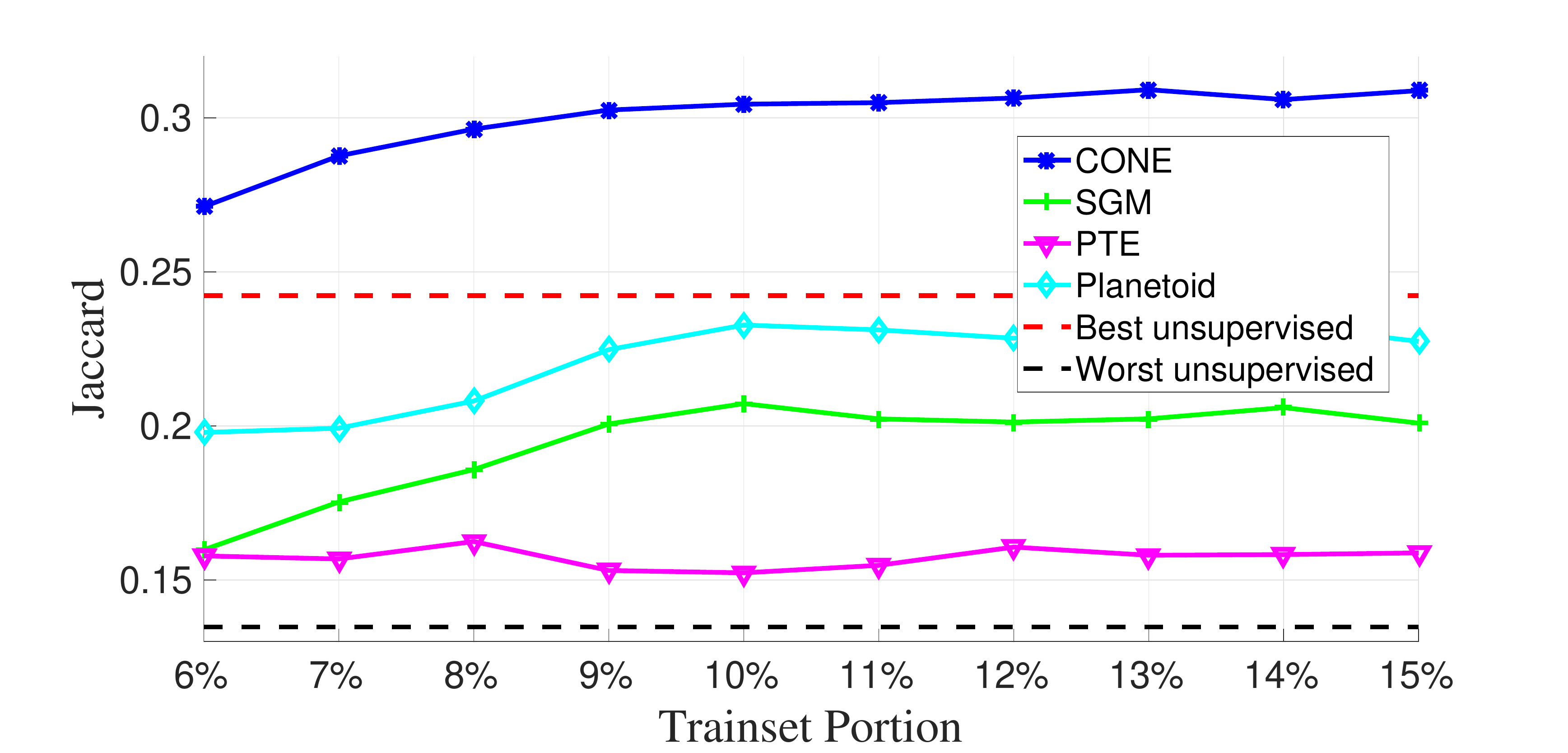}}
\hspace{-15pt}
\subfigure[Flickr]{
\includegraphics[width=0.335\textwidth]{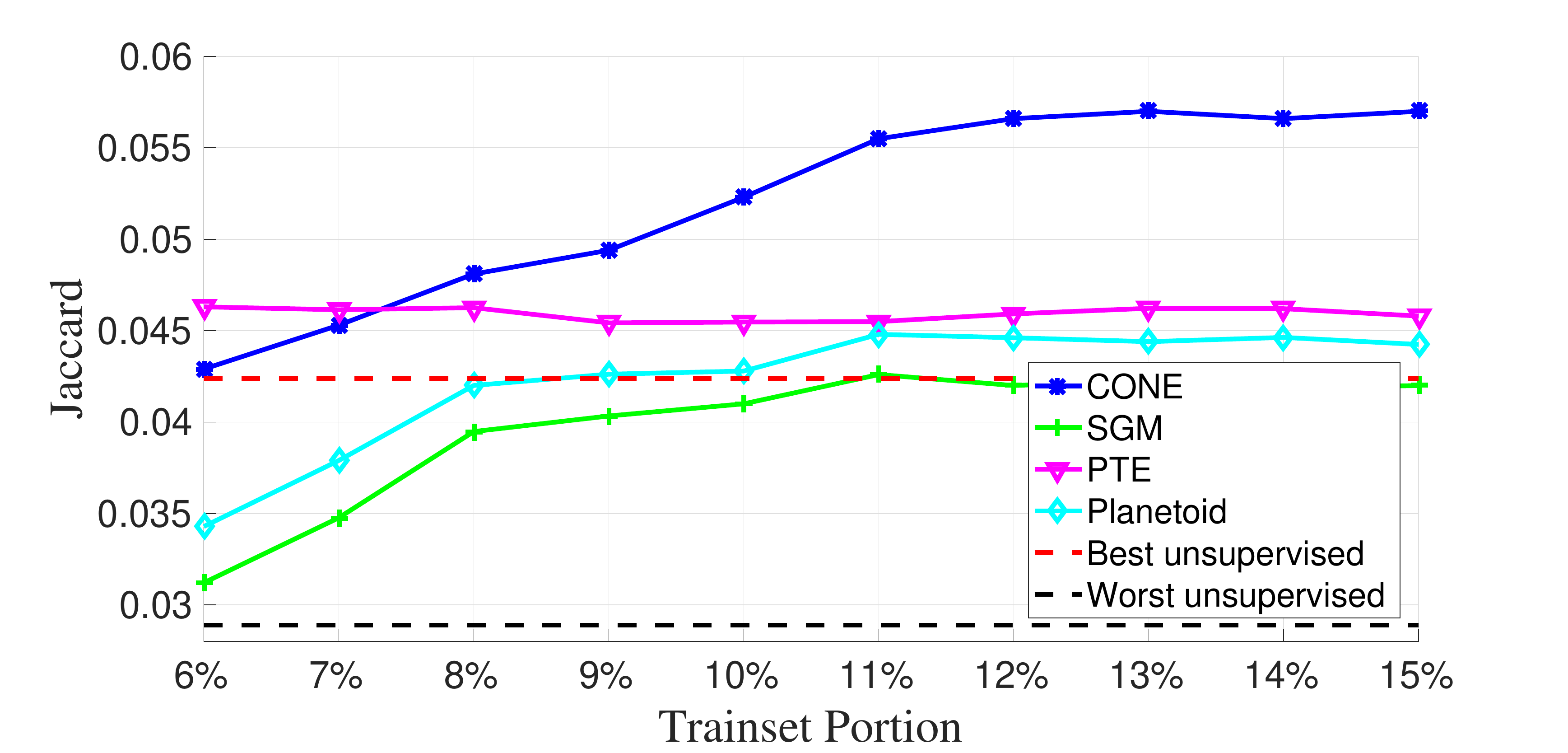}}
\hspace{-15pt}
\subfigure[Twitter]{
\includegraphics[width=0.335\textwidth]{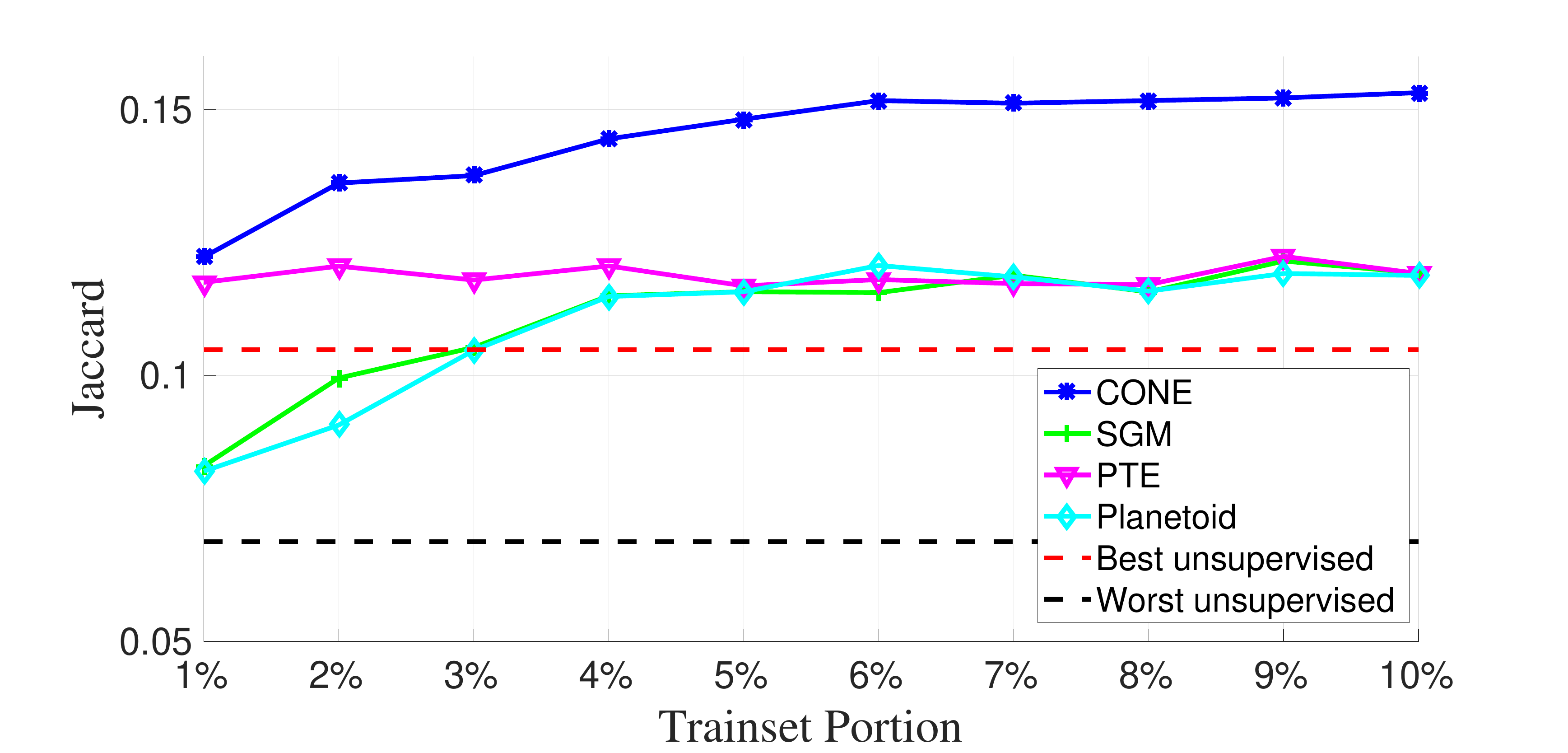}}
\caption{\textbf{Model selection results with varying training set portion.}}
\label{fig:trainset}
\end{figure*}

{\flushleft \bf Impact of supervision.} 
We study the impact of supervision on the performance of CONE by varying the training and testing set portions. Figure \ref{fig:trainset} shows the results. CONE's advantage of leveraging community examples is significant on all three datasets. As can be observed, CONE efficiently leverages supervision by learning from very small amounts of labeled communities, and the performance converges rapidly as the training set portion reaches around 11\% on the Facabook and Flickr datasets. On datasets with large amounts of labeled communities such as Twitter, around 6\% of them are enough to learn a good CONE model. 

Note that the other three supervised algorithms do not leverage community labels effectively, basically because they are not designed to capture social patterns and leverage them for out-of-sample community detection.

{\flushleft \bf Impact of architecture.} 
We study the impact of parameters inside CONE by varying the number of random-walk transitions and the size of embeddings. We also substitute the RNN in the content embedding layer with basic fully connected feed forward neural networks to demonstrate the effectiveness of using RNN to deeply explore the high-dimensional noisy contents. Among the basic neural networks we experimented on with varying layers (from 1 to 4) and embedding sizes (the last layer with sizes of 4, 8, 16 and 32), we show the performance of a three-layer basic neural network (embedding sizes are $64\to32\to16$ from bottom up as a common practice \cite{He2017Neural}) with ReLU as the activation function, which generally yields the best performance. 

\begin{figure*}[h!]
\centering
\subfigure[Facebook]{
\includegraphics[width=0.335\textwidth]{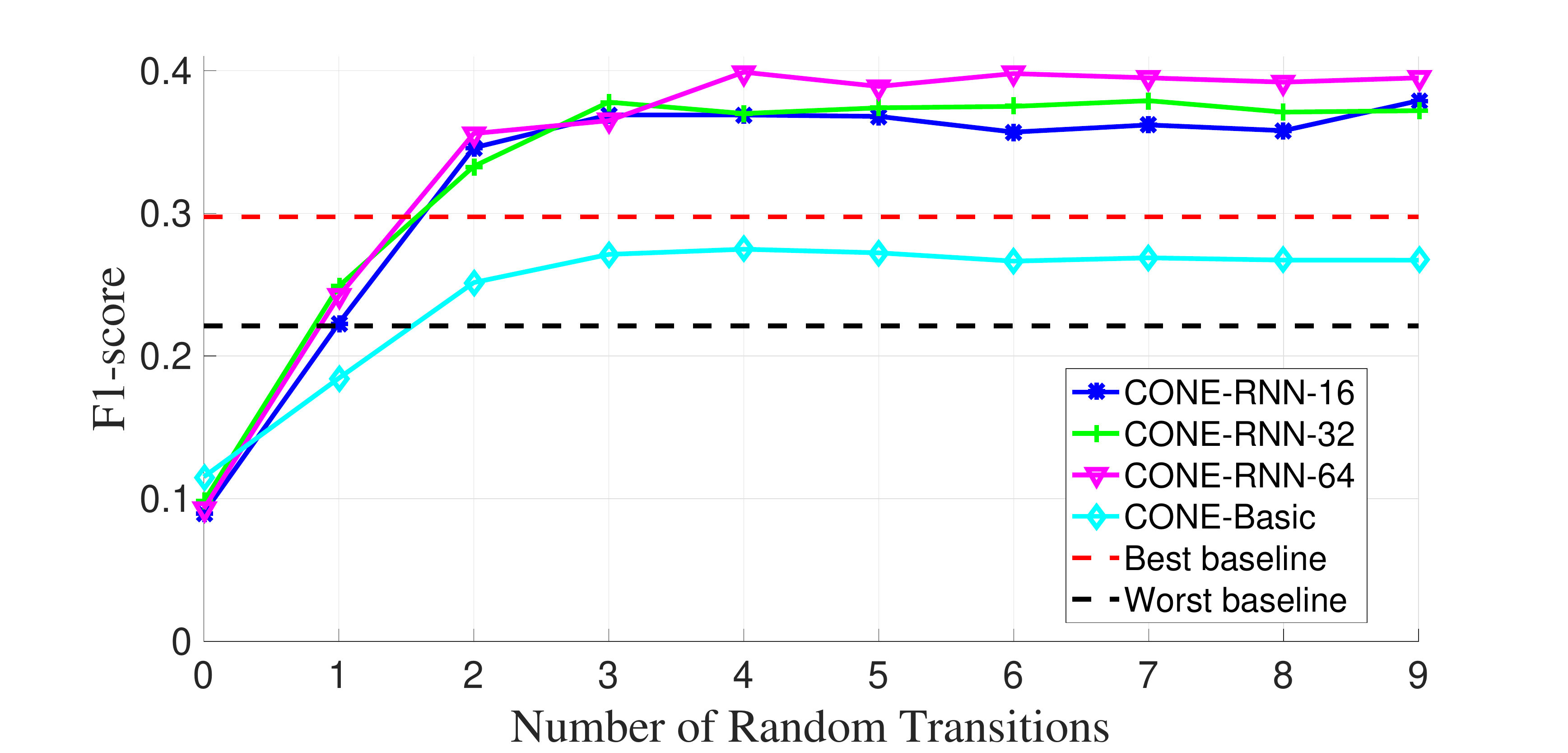}}
\hspace{-15pt}
\subfigure[Flickr]{
\includegraphics[width=0.335\textwidth]{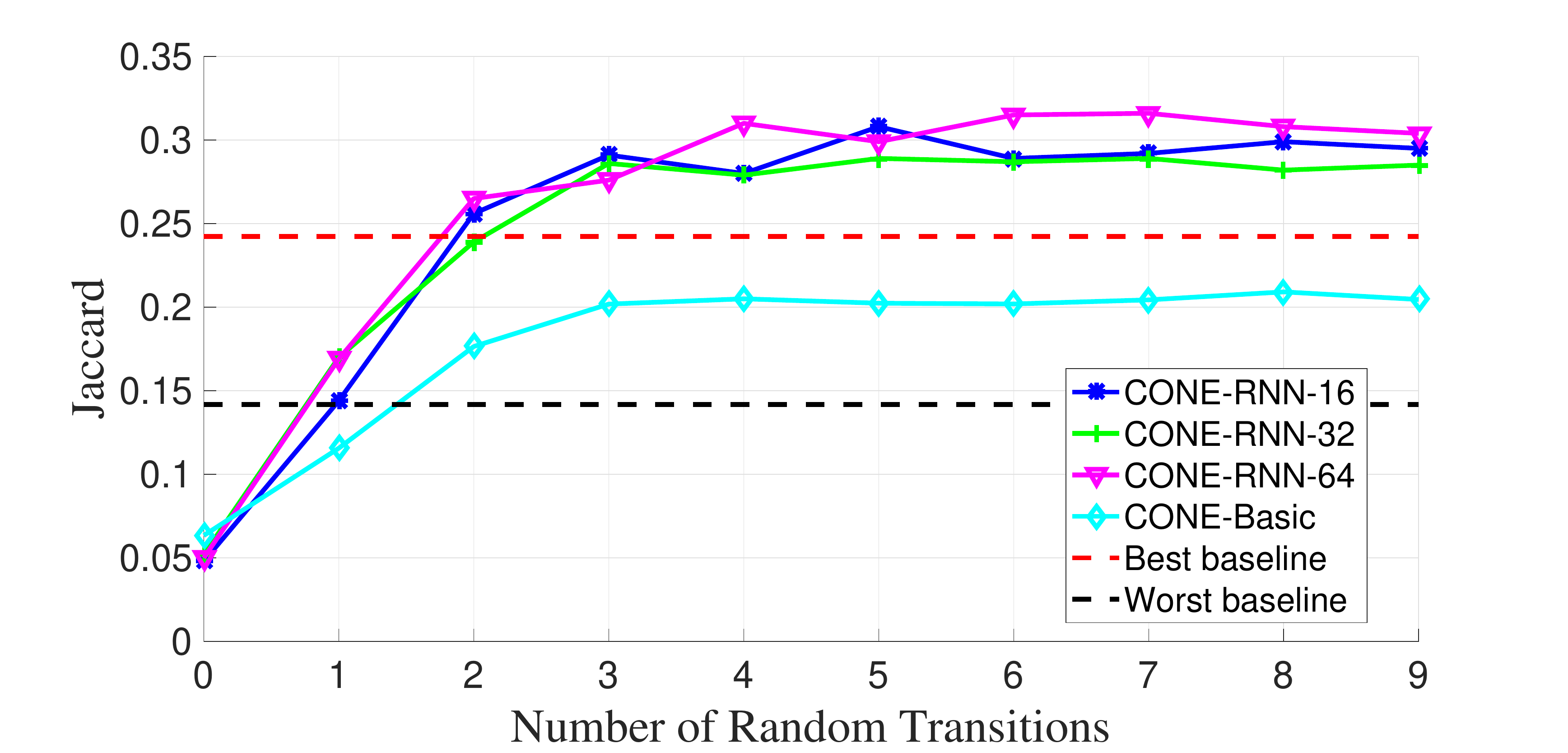}}
\hspace{-15pt}
\subfigure[Twitter]{
\includegraphics[width=0.335\textwidth]{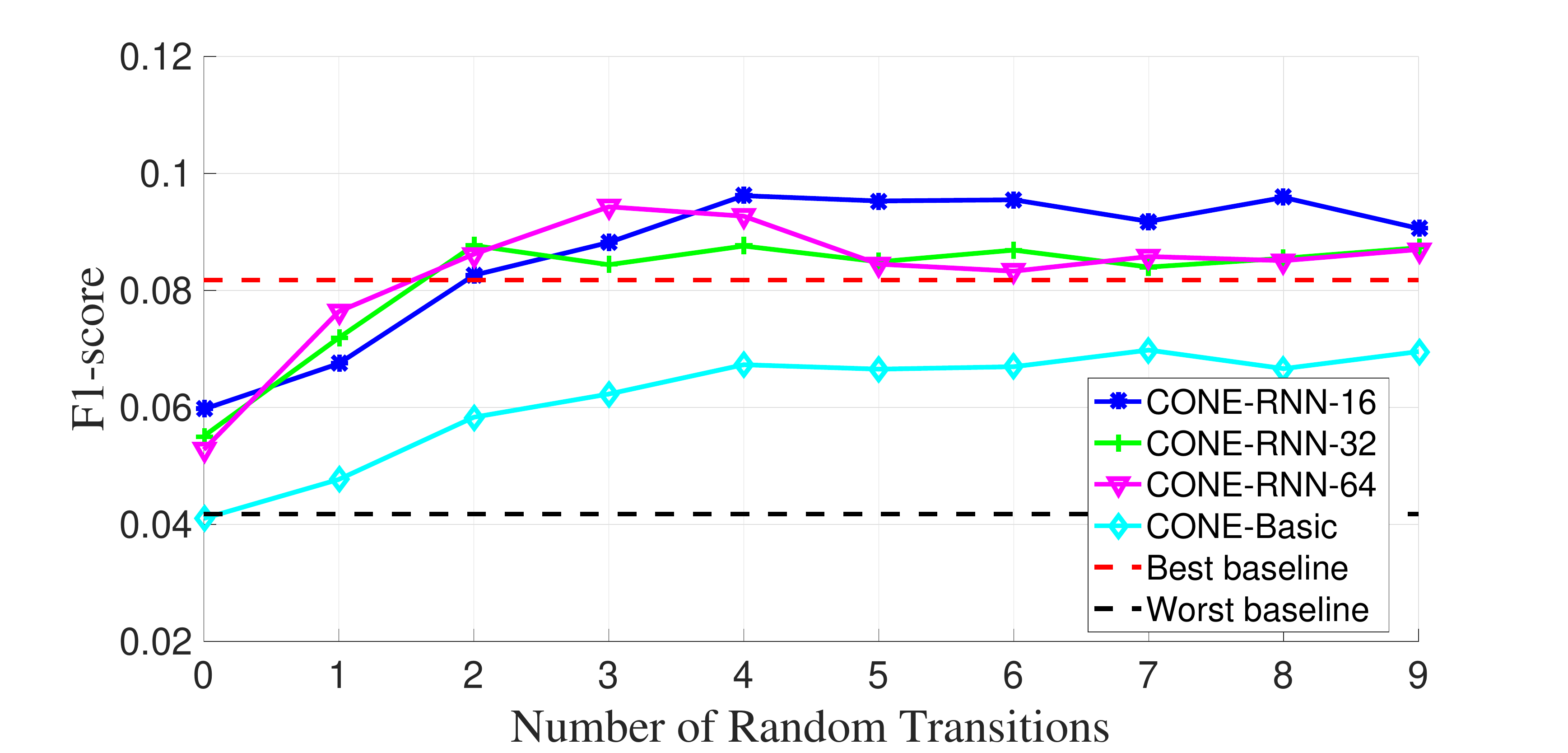}}
\subfigure[Facebook]{
\includegraphics[width=0.335\textwidth]{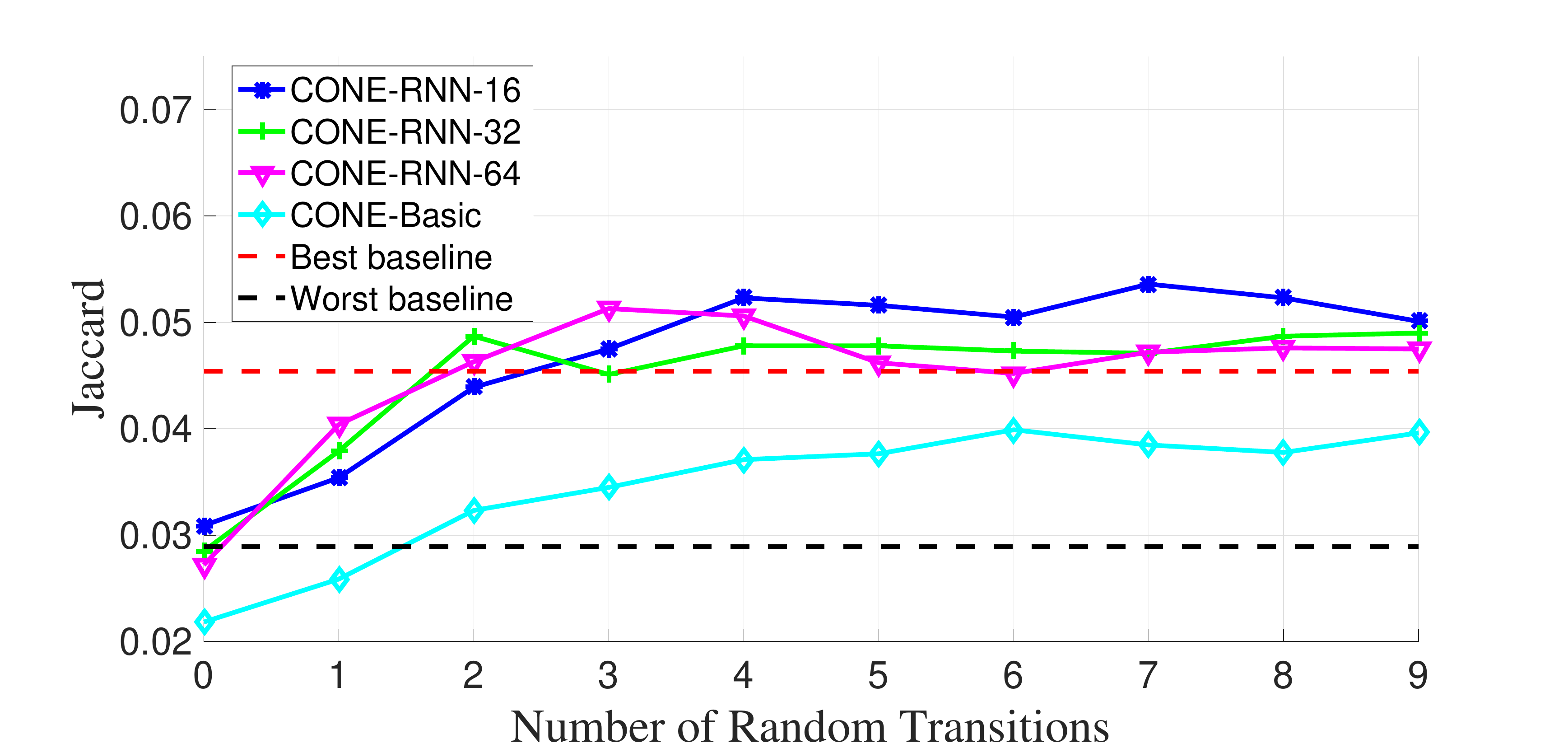}}
\hspace{-15pt}
\subfigure[Flickr]{
\includegraphics[width=0.335\textwidth]{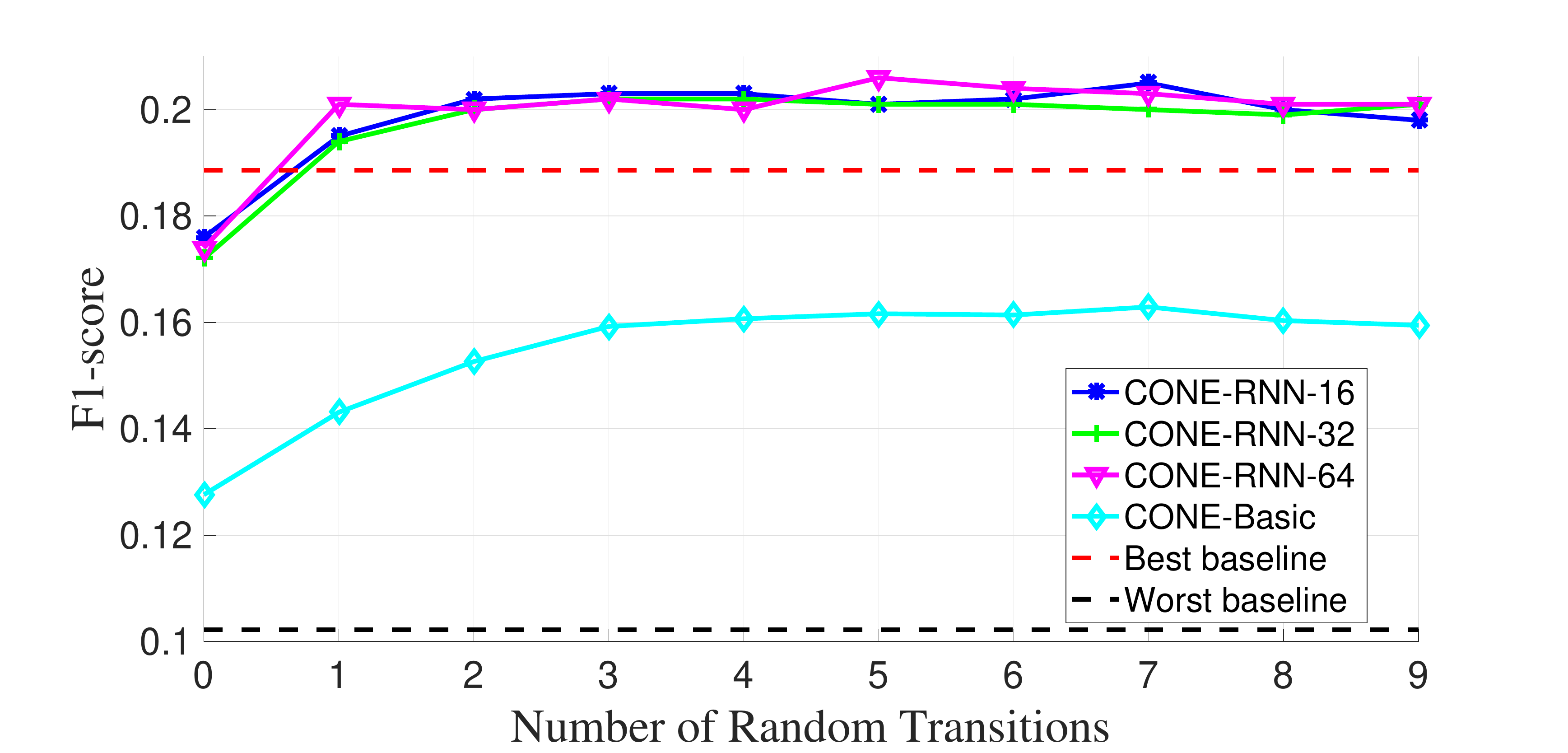}}
\hspace{-15pt}
\subfigure[Twitter]{
\includegraphics[width=0.335\textwidth]{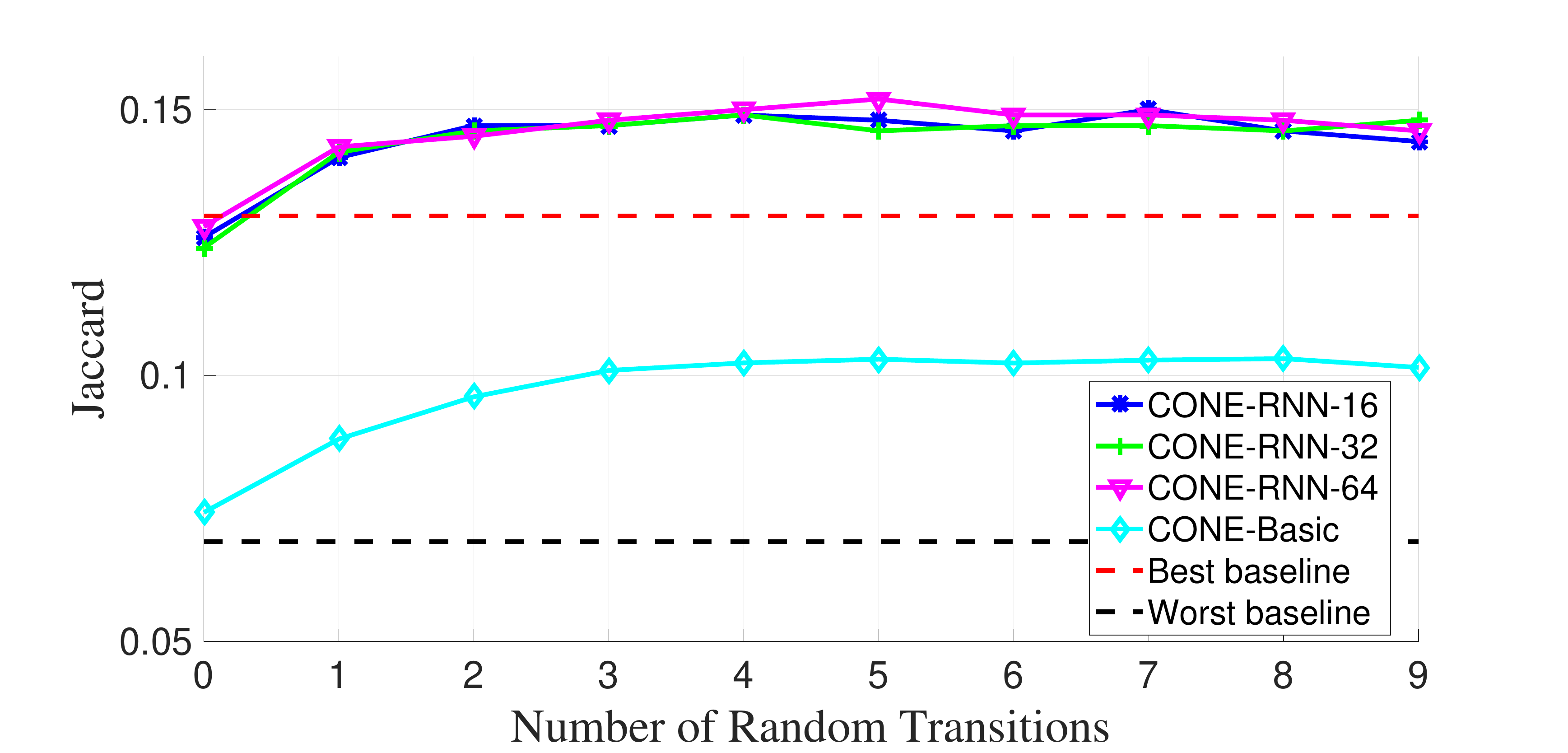}}
\caption{\textbf{Model selection results with varying neural architecture.}}
\label{fig:architecture}
\end{figure*}

Figure \ref{fig:architecture} shows the results. The number of random-walk transitions has a large impact on the performance of CONE, especially when the number is small. This demonstrates the utility of our novel network reconfiguration layer. As the number of transitions grows larger and the stationary distribution of random walk is approximated, the performance becomes stable. Note that large numbers of transitions do not necessarily lead to optimal performance, because community structures are often well described within small neighborhoods. 

The number of embedding dimensions does not have a significant impact on the performance with RNN as the content embedding layer, indicating the robustness of CONE inherited from RNN. However, the performance is significantly worse without RNN as the content embedding layer, which justifies our motivation of leveraging RNN to effectively explore the noisy contents in high-dimensional spaces. The runtime of RNN is also significantly shorter than basic neural networks, especially with more layers and larger embedding sizes.

\subsection{Case Study}
\begin{figure}[h!]
\centering
\subfigure[Density in ego-net 686]{
\includegraphics[width=0.23\textwidth]{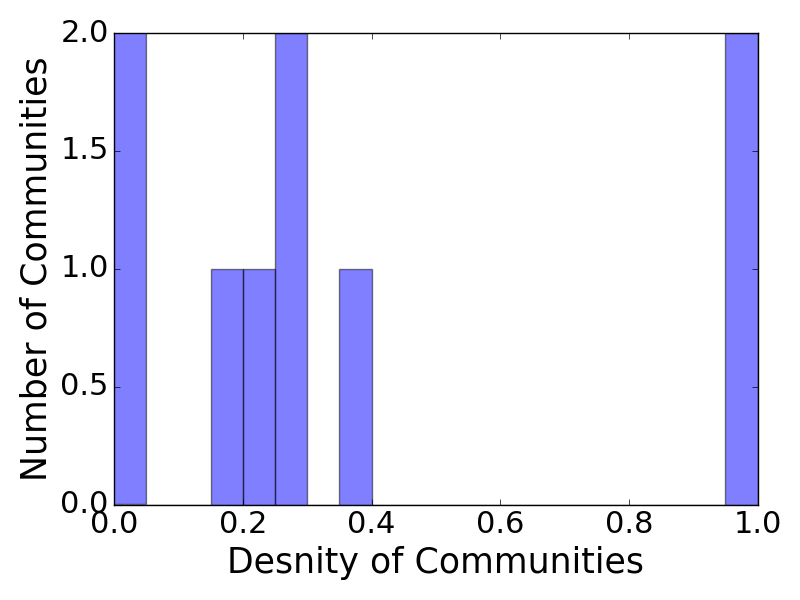}}
\subfigure[\small Homogeneity in ego-net 698]{
\includegraphics[width=0.23\textwidth]{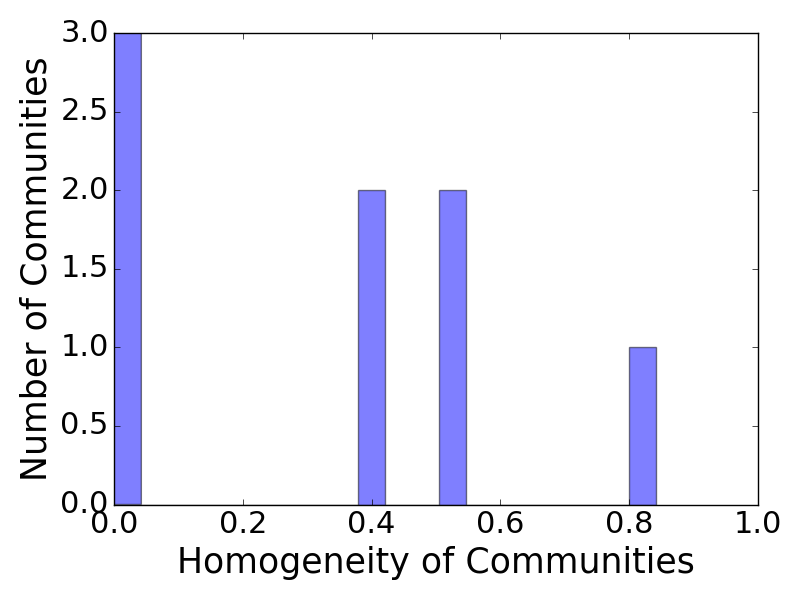}}
\caption{\textbf{Communities detected by CONE.}}
\label{fig:case}
\end{figure}

To understand the advantage of CONE, we conduct the same density and homogeneity analysis on our detected communities and present the results on the same ego-networks we showed in Sec.~\ref{sec:analysis}. Comparing Figure \ref{fig:case}(a) and \ref{fig:case}(b) with Figure \ref{fig:dense}(a) and \ref{fig:homo}(a), we observe that CONE indeed finds communities that are similar in distributions of density and homogeneity as the ground-truth, which indicates its effectiveness in capturing the social patterns that essentially define user-described communities.

We also looked into the structures of communities found by different algorithms. Still taking the example of the clear co-star pattern, when we took the ground-truth communities in ego-network 1684 as supervision, CONE indeed found a total of 6 co-star communities on other parts of the dataset, specifically, in ego-networks 0, 107, 348 and 3437. We did not observe communities of such a co-star pattern detected by any other algorithms, from all communities they detected through a single execution on the same training and testing sets.

\section{Related Work}

\subsection{Community detection}
Traditional community detection algorithms are mostly unsupervised, based on either of the edge density or node homogeneity assumptions, or both. Therefore, the communities they detect are only constructed by densely connected or homogeneous nodes.

Among them, many are based on probabilistic generative models. 
For instance, COCOMP \cite{zhou2012community} extends LDA \cite{blei2003latent} by taking community assignments as latent variables. It leverages node homogeneity by assuming that each community is a group of people that use a specific distribution of words. BigClam \cite{yang2013overlapping} does not make use of node content at all. It leverages edge density by modeling the assignments of communities solely based on existing links. A variation of BigClam called CESNA \cite{yang2013community} is later proposed, with additional consideration of binary-valued node attributes. 
Another method within the state-of-the-art, Circles \cite{mcauley2012learning}, also leverages the same assumptions with a slightly different model. 

Many non-generative algorithms have also been proposed. Among them, some are based on graph metrics such as modularity \cite{clauset2004finding} and maximal clique \cite{du2007community}. They are fundamentally leveraging the edge density assumption. As for node homogeneity, some algorithms firstly augment the graph with content links and then do clustering on the augmented graph \cite{ruan2013efficient,yang2009combining}, while some others attempt to find a partition that yields the minimum encoding cost based on information theory \cite{akoglu2012pics,rosvall2008maps}. They do not scale trivially to networks with high-dimensional noisy contents.

As we leverage whole labeled communities as examples to learn the underlying social patterns, CONE is also different from the few semi-supervised community detection algorithms that leverage individual labels and pair-wise constraints as part of some communities \cite{eaton2012spin}.

\subsection{Network embedding}
We aim to find node embeddings that are suitable for the task of community detection, under the social pattern assumption. Unlike the techniques that compute embeddings based on node proximities in the network \cite{henderson2011s, tang2009relational, tang2011leveraging}, we frame the objective as learning a representation that captures social patterns. Moreover, we aim to learn it under the guidance of communities, instead of the unsupervised ways based on relational data in the networks \cite{ahmed2013distributed, yan2007graph} or structural relationships between concepts \cite{dahl2012context, krizhevsky2012imagenet}. 

As discussed in Sec.~\ref{sec:model}, the most related techniques to ours are the Skipgram-based network embeddings \cite{grover2016node2vec,perozzi2014deepwalk,tang2015line}. Recent works \cite{tang2015pte,yang2016revisiting,yang2015network} have extended such techniques to incorporate node attributes and class labels. Their models using augmented nodes, text feature matrices and bag-of-word vectors are ineffective for deeply exploring high-dimensional noisy contents.
Moreover, they all use network context as supervision to uniformly require nodes with similar context to have similar embeddings. Our objective is different, where we only require users within the same communities to be embedded as close. To this end, our supervision is directly guided by community examples, and we leverage network structures only as a regularization.
Finally, since they sample networks into paths and predict context based on paths, the techniques are inefficient in capturing local network structures beyond paths. Our regularization efficiently generalizes the embedding of single users to their ambient local structures, which are deterministic rather than approximated by paths.

\subsection{Node classification}
Traditional supervised learning on networks is essentially node classification \cite{bhagat2011node,he2004manifold,tang2009relational,zhu2003semi}, which uses node attributes as labels and leverages network structures for tasks like content prediction. Our problem is quite different from node classification. Although we utilize supervision to learn the community oriented features, our final goal is to perform unsupervised detection of unlabeled communities. To be more specific, our predictions are new clusters of nodes, for which no predefined categories or labeled data are available at all. As a consequence, node classification algorithms are not applicable to our problem.
\section{Conclusion}
In this paper, we doubt the generality of the two widely trusted assumptions about community characteristics, \ie, edge density and node homogeneity, and we show the deficiency of mainstream algorithms adopting those assumptions through real data analysis. To deal with this deficiency, we propose to leverage the underlying social patterns that define and detect network communities. We design CONE that effectively explores and captures important social patterns under the guidance of example communities through network embedding. Generic clustering algorithms performed on the embeddings can yield reliable community detection results.


While CONE was originally designed for leveraging supervision for reliable community detection free from falsifiable assumptions, it can be easily applied to many network learning tasks to coherently leverage node contents, link structures and various kinds of supervision or constraints. As we observe from our experiments, CONE is especially advantageous in dealing with high-dimensional noisy contents, such as sequences of hashtags and even raw texts.  
\bibliographystyle{IEEEtran}
\small
\bibliography{carlyang} 
\end{document}